%% file: main.tex
\documentclass[11pt,a4paper]{article}
\pdfoutput=1

\usepackage{jheppub}
\usepackage[utf8]{inputenc}
\usepackage{xspace}

\usepackage{verbatim}
\usepackage{amsthm,mathtools}
\usepackage{relsize}
\usepackage[mathscr]{euscript}
\usepackage[cal=dutchcal]{mathalfa}
\usepackage{scalerel}
\pdfmapfile{+rsfso.map}
\DeclareSymbolFont{rsfso}{U}{rsfso}{m}{n}
\DeclareSymbolFontAlphabet{\mathscr}{rsfso}

\usepackage{float}
\usepackage[section]{placeins}
\usepackage[lofdepth,lotdepth]{subfig}
\usepackage[toc,page]{appendix}
\usepackage[dvipsnames]{xcolor}
\usepackage{xtab}
\usepackage{enumitem}
\usepackage[hhmmss]{datetime}

\usepackage{tikz}
\usetikzlibrary{shapes,arrows}

\usepackage{booktabs}
\allowdisplaybreaks

\theoremstyle{definition}

\theoremstyle{definition}

\theoremstyle{plain}

\theoremstyle{plain}

\theoremstyle{remark}

\theoremstyle{remark}

\usepackage{tabu}
\usepackage{booktabs}
\usepackage{multirow}
\newcolumntype{P}[1]{>{\centering\arraybackslash}p{#1}}

\hypersetup{hidelinks}

\makeatletter

\AtBeginDocument{%
	\expandafter\renewcommand\expandafter\subsection\expandafter{%
		\expandafter\@fb@secFB\subsection
	}%
}

\makeatother % this adds \FloatBarrier at the end of subsections, so floating objects are obliged to stay in the subsection you want them to stay

\newcommand{\rb}[1]{\left(#1\right)}
\newcommand{\qb}[1]{\left[#1\right]}
\newcommand{\cb}[1]{\left\lbrace #1 \right\rbrace}

\newcommand{\mydet}[1]{\operatorname{det}\left({#1}\right)}

\newif\ifColors    \Colorsfalse
\newif\ifShowSigns    \ShowSignstrue

\hypersetup{
	pdfstartview={Fit}, pdfpagelayout={SinglePage}
}

\input{macros.tex}

\title{The mean gauges in bimetric relativity}

\author{Francesco Torsello}
\emailAdd{francesco.torsello@fysik.su.se}

\affiliation{Department of Physics \& The Oskar Klein Centre, \\
Stockholm University, AlbaNova University Center, SE-106 91 Stockholm, Sweden}

\abstract{The choice of gauge in numerical relativity is crucial in avoiding coordinate and curvature singularities. In addition, the gauge can affect the well-posedness of the system. In this work, we consider the mean gauges, established with respect to the geometric mean metric $\hMet \coloneqq \gMet \, \bigrb{\gMet^{-1}\fMet}^{1/2}$ in bimetric relativity. We consider three gauge conditions widely used in numerical relativity, and compute them with respect to the geometric mean: The \onePlusLog gauge condition and the maximal slicing for the lapse function of $\hMet$, and the $\Gamma$-driver gauge condition for the shift vector of $\hMet$. In addition, in the bimetric covariant BSSN formalism, there are other arbitrary choices to be made before evolving the system. We show that it is possible to make them by using the geometric mean metric, which is determined dynamically by the system, rather than using an arbitrary external metric, as in general relativity. These choices represent opportunities to recast the system in a well-posed form.}

\keywords{ghost-free bimetric theory, Hassan--Rosen bimetric theory, bimetric relativity, standard gauge, maximal slicing, geometric mean, numerical relativity.}

\newcommand{\change}[1]{{#1}\xspace}

\begin{document}

\maketitle

\section{Introduction and background}
\label{sec:introduction}

The Hassan--Rosen bimetric theory, or bimetric relativity (BR) \cite{Hassan:2011zd,Hassan:2011ea,HassanKocic2018,HassanLundkvist2018} is a theory of two nonlinearly interacting \change{Lorentzian} metrics $\gMet$ and $\fMet$ defined on the same differentiable manifold. \change{The theory is free from the Boulware--Deser ghost \cite{PhysRevD.6.3368} and propagates seven degrees of freedom which, for linear perturbations around backgrounds with proportional metrics, can be recognized as the polarizations of a massless and a massive spin-2 field, expressed as linear combinations of the two metric fields \cite{PhysRevD.85.024044,Hassan2013,Hassan_2013}. The introduction of the second metric allows for a description of a massive spin-2 field \cite{PhysRevD.3.867}.}

\change{The study of this theory is interesting in many respects. First, it is the only consistent classical field theory of two nonlinearly interacting spin-2 fields to date, therefore it is a promising source of novel physics. We remind the reader that there cannot be interaction between massless spin-2 fields \cite{BOULANGER2001127}, hence the existence of a consistent theory of interacting massless and massive spin-2 fields is remarkable.}

\change{Second, it is a theory where two Lorentzian metrics are coupled already off-shell, that is, before solving the field equations of the theory, and this has intriguing geometrical consequences. In more detail, the Boulware--Deser ghost is avoided by a cautious choice of the interaction potential \cite{PhysRevD.82.044020,PhysRevLett.106.231101}, defined in terms of the principal square root $\mSqrt\coloneqq \rb{\gMet^{-1}\fMet}^{1/2}$ of the $(1,1)$ tensor field $\gMet^{-1}\fMet$ \cite{HassanKocic2018}. For instance, in geometrized units where $c=G=1$, the action reads,}
\begin{align}
\label{eq:br-action}
	\eulS	&\coloneqq \int\dd^4 x\sqrt{-g} \biggqb{\biggrb{\dfrac{1}{2\gKappa}\gRicci + \gLmat} +\bimV(\mSqrt) + \mydet{\mSqrt}\biggrb{\dfrac{1}{2\fKappa}\fRicci + \fLmat}},
\end{align}
\change{where $\gKappa$ and $\fKappa$ are the dimensionless Einstein's gravitational constants, $\gLmat$ and $\fLmat$ are independent matter Lagrangians \cite{PhysRevD.90.124042,Rham_2015} and the bimetric potential is}
\begin{align}
\label{eq:bim-V}
	\bimV(\mSqrt)	&\coloneqq -\dfrac{1}{\ell^2}\,\sum_{n=0}^4\beta_{(n)}e_n(\mSqrt),
\end{align}
\change{with $\ell $ length scale of the interaction, $\beta_{(n)}$ dimensionless real free parameters and $e_n(\mSqrt)$ elementary symmetric polynomials of $\mSqrt$. Hence, in order to be able to define the bimetric action \eqref{eq:br-action}, the invertible real square root $\mSqrt$ must exist. The existence of $\mSqrt$ is equivalent to setting profound geometrical constraints on the two metrics $\gMet$ and $\fMet$, namely that they are causally or null coupled \cite{HassanKocic2018}, that is, either a hypersurface which is spacelike with respect to both metrics and a direction which is timelike with respect to both metrics exist, or none of them exist. Therefore, the theory also provides a fertile ground to study Lorentzian geometry in the presence of several metrics.}

\change{Finally, BR is a geometric theory of gravity where the gravitational origin of dark matter can be explored \cite{Aoki:2014cla,Blanchet:2015bia,Enander:2015kda,Babichev:2016bxi}, and whose spectrum of solutions includes self-accelerating cosmologies \cite{Volkov:2011an,vonStrauss:2011mq,Comelli:2011zm,Akrami:2012vf}. In addition, BR is compatible with the observation of gravitational waves and with local gravity tests \cite{10.1088/1361-6382/ab4f9b}.}

\change{Due to all of these reasons,} it is desirable to obtain solutions describing realistic systems within this theory, e.g., spherical gravitational collapse of matter, or gravitational waves emitted by interesting physical systems. This would lead to a direct comparison with the observational data and would make it possible to falsify or support the theory. In order to solve for solutions describing realistic physical systems, it is necessary to recast the bimetric field equations (BFE) in a form suitable for numerical integration. In this respect, following the roadmap outlined by numerical relativity, the standard \nPlusOne decomposition of the BFE governing the dynamics of the two metrics was established in \cite{Kocic:2018ddp}, and the covariant BSSN (cBSSN) formulation \cite{PhysRevD.79.104029} of the bimetric decomposition was presented in \cite{Torsello_2019}. However, the bimetric cBSSN formulation together with the standard gauge \cite{PhysRevLett.75.600,PhysRevD.67.084023} relative to one of the metrics, was not proven to be strongly hyperbolic under the same assumptions as in general relativity (GR). This is due to the fact that the lapse functions of the metrics are dependent \cite{HassanLundkvist2018,Alexandrov2013}. For instance, their ratio $\mW$ is a rather complicated function of the dynamical fields and their spatial derivatives, and presently its explicit form is known only in spherical symmetry \cite{Kocic_2019}. The hyperbolic structure of the system is altered compared with GR (even after a first-order reduction) by the spatial derivatives of $\mW$, which contain the spatial derivatives of the dynamical fields.

Therefore, it is fundamental to study other gauge choices in BR, which could potentially lead to a well-posed formulation of the bimetric cBSSN formulation. In this respect, one novel feature of BR with respect to GR is the presence of the geometric mean sector. The geometric mean metric $\hMet$ of $\gMet$ and $\fMet$ is defined as \cite{Hassan2013,HassanKocic2018},
\begin{align}
	\hMet \coloneqq \gMet \operatorname{\#} \fMet = \gMet \rb{\gMet^{-1}\fMet}^{1/2}=\gMet\, \mSqrt.
\end{align}
The lapse function $\hLapse$ and the shift vector $\hShift$ of $\hMet$ are related to those of $\gMet$ and $\fMet$, hence we can gauge fix $\hLapse$ and $\hShift$, and after this choice the lapses and shifts of $\gMet$ and $\fMet$ will also be determined \cite{Kocic:2018ddp}, as summarized in \autoref{subsec:gaugefixing}.

In this paper, we study several gauge conditions with respect to the geometric mean metric $\hMet$, the \qm{mean gauges}. This uncharted territory is explored for three main reasons:
\begin{enumerate}
	\item To get insights about how to recast the cBSSN equations in a well-posed form, by choosing a suitable gauge condition.
	\item To understand if it is possible to impose a gauge condition which is singularity avoiding for \emph{both} $\gMet$ and $\fMet$. Since $\hMet$ is their geometric mean, and its null-cone is always contained in the convex hull of the null-cones of $\gMet$ and $\fMet$ \cite{HassanKocic2018}, the hope is that imposing a singularity avoiding condition on $\hMet$ also avoids singularities in $\gMet$ and $\fMet$.
	\item To find new gauges, specific to bimetric relativity, which may lead to a stable long-term numerical simulations, as outlined in \cite{10.1143/PTP.101.251} for general relativity.
\end{enumerate}
Hence, this study is motivated both theoretically and numerically. Indeed, since the computation of $\mW$ is a source of several numerical errors, the BFE are plagued by more sources of instability compared with the Einstein field equations (EFE); therefore, well-posedness is a fundamental requirement. There are also some hints suggesting the possibility to find a singularity avoiding gauge for both metrics simultaneously. For example, Proposition 1 in \cite{PhysRevD.96.064003} establishes that $\mydet{\mSqrt}=0$ at a point in spacetime induces a curvature singularity in one of the two metrics $\gMet$ and $\fMet$, if the other is regular at that point. This result suggests that, when one of the three metrics $\gMet,\fMet,\hMet$ has a curvature singularity at a point in spacetime, one of the other two metrics should share it. Supporting this conjecture, an exact solution showing a curvature singularity in $\fMet$ and $\hMet$, but not in $\gMet$, can be found in \cite{Hogas:2019cpg}. In addition, in \cite{PhysRevD.97.084022} it is pointed out that, since the metrics are defined on the same differentiable manifold, they must be compatible with its topology, which is fixed by the choice of an atlas. Since the presence of a singularity usually changes the topology of the manifold, and the metrics must be compatible with the same topology, the hint is again that the metrics should share the singularity (which, however, does not need to be of the same nature for all the metrics). However, as we point out in the paper, finding a singularity avoiding gauge for both metrics is nontrivial. For example, imposing maximal slicing simultaneously for both of the metrics results in a system of two partial differential equations for one of the lapse functions, and one has to prove that a solution to this system exists. This motivates further the study of the mean gauges.

In this paper, we compute the standard gauge and the maximal slicing with respect to $\hMet$, and discuss some of their features. We do not study how these gauges affect the hyperbolicity of the system, which is an open problem whose solution is beyond the scope of this work. The reason relies on the fact that the explicit expression for $\mW$, shown in \cite{Kocic_2019} in the standard \threePlusOne formulation and in the ancillary files of \cite{Torsello_2019} in the cBSSN formulation, is not easy to deal with. The study of the hyperbolicity of the system involves the first and second spatial derivatives of $\mW$, hence it is left for future work. In addition, as a natural continuation of the work done in \cite{Torsello_2019}, we discuss the possibility to use the geometric mean sector as the background sector for the conformal spatial metrics $\gSpBS,\fSpBS$, and to use it to fix the evolution of the determinant of the conformal metrics in the bimetric cBSSN formalism. These are novel opportunities in BR with respect to GR, to recast the system in a well-posed form. A brief review on the background geometries in cBSSN is given in \autoref{subsec:cBSSNobservers}.

\paragraph{Structure of the paper.} In \autoref{subsec:gaugefixing} we provide more details about gauge fixing in BR, and in \autoref{subsec:cBSSNobservers} we introduce the concept of background geometries in the cBSSN formalism. We establish the dynamics of the spatial part of the geometric mean in \autoref{sec:geometricmean}. We present the mean standard gauge and the mean maximal slicing in \autoref{sec:meangauges}, and discuss the usage of the geometric mean in the cBSSN formalism in \autoref{sec:meanBSSN}. Finally, we state our conclusions in \autoref{sec:conclusions}. The appendices contain the notation and technical details.

\paragraph{Notation.} We follow the same notation as in \cite{Torsello_2019}, described in Appendix~\ref{app:notation}.

\subsection{Gauge fixing in bimetric relativity}
\label{subsec:gaugefixing}

In the \nPlusOne decomposition of BR, the mean lapse function $\hLapse$ and the mean shift vector $\hShift$ of $\hMet$ are related to those of $\gMet$ and $\fMet$ by \cite{Kocic:2018ddp},
\begin{subequations}
\label{eq:gaugevariables}
	\begin{alignat}{3}
		\gLapse^2 &=\hLapse ^2 \sLt\mW, \qquad & \fLapse ^2 &= \dfrac{\hLapse ^2 \sLt}{\mW}, \qquad & \gLapse &=\mW \fLapse ,\label{eq:gaugevariableslapses} \\
		\gShift &=\hShift + \dfrac{\gLapse}{\sLt} \gE^{-1}\sLp , \qquad & \fShift &=\hShift - \dfrac{\fLapse}{\sLt} \fE^{-1}\sLp, \qquad & \fShift - \gShift &= -\dfrac{\fLapse}{\sLt}\rb{ \mW\,\gE^{-1}+\fE^{-1} }\sLp.\label{eq:gaugevariablesshifts}
	\end{alignat}
\end{subequations}
Here, $\gLapse$ and $\gShift$ are the lapse function and shift vector of $\gMet$, whereas $\fLapse$ and $\fShift$ are the lapse function and shift vector of $\fMet$, $\sLp$ is a Lorentz spatial vector called \qm{separation parameter} since it parameterizes the difference between the shifts, and $\sLt\coloneqq\rb{1+\sLp^\tr \sEta \sLp}^{1/2}$, with $\sEta$ being the spatial part of the Minkowski metric in the Lorentz frame (i.e., the three-dimensional Euclidean metric). Also, $\gE$ and $\fE$ are the spatial vielbeins of the spatial parts $\gSp, \fSp$ of the metrics $\gMet,\fMet$. The lapses and shifts altogether are called the \qm{gauge variables.}

We have the freedom to gauge fix one of the three lapse functions and one of the three shift vectors (or one linear combination of the lapses and one linear combination of the shifts) thanks to diffeomorphism invariance. After this gauge fixing, since $\sLp$ and $\mW$ in \eqref{eq:gaugevariables} are determined by the dynamical fields, the other gauge variables will also be fixed, as it was already mentioned in \cite{Kocic:2018ddp,Kocic_2019,Torsello_2019}. At this point it is helpful to introduce some terminology. We say that we \qm{choose a gauge condition with respect to a metric}, to mean that the slicing of spacetime is done with respect to that metric. For example, imposing the maximal slicing with respect to $\hMet$ means that we impose the gauge condition
\begin{align}
\label{eq:maxslicexample}
\hK=\dt\hK=0,
\end{align}
where the accent $\#$ means that the quantity refers to the geometric mean sector (see Appendix \ref{app:notation}). We can solve \eqref{eq:maxslicexample} for either one of $\hLapse,\fLapse,\gLapse$. Therefore, the same gauge choice with respect to a metric, can be solved for any gauge variable. We say that we \qm{gauge fix} the variable which we solve the gauge condition for. Different gauge fixings that refer to the same gauge condition are equivalent geometrically, but not necessarily analytically since the equations for different gauge variables are different. Hence, the well-posedness of the system may be affected not only by the geometric gauge condition, but also by the choice of the gauge fixed variable.

\subsection{The background geometries in the covariant BSSN formalism}
\label{subsec:cBSSNobservers}

The cBSSN formulation of the EFE \cite{PhysRevD.79.104029} introduces the background connection $\gGbackBS^i_{jk}$ \cite{Garfinkle_2008}, needed to make one of the dynamical variables, namely the conformal connection $\gL^i$, a vector under spatial coordinate transformations not involving the time coordinate. In GR this background connection can, but need not, be the Levi--Civita connection of a background metric. In \cite{PhysRevD.79.104029}, the background connection is restricted to not depend on time, since it is arbitrary and it is not constrained by any physically motivated evolution equation.

In \cite{Torsello_2019}, the possibility of relaxing the restriction of time-independence of the background connection in BR is mentioned. Indeed, in BR we can set the conformal spatial part $\hSpBS$ of the geometric mean $\hMet$ to be the background metric defining the background connection for both of the conformal metrics $\gSpBS$ and $\fSpBS$. The dynamics of the conformal spatial part $\hSpBS$ of $\hMet$ can be computed in terms of the dynamics of $\gSpBS$, $\fSpBS$ and the separation parameter $\sLp$. This allows us to relax the assumption of time-independence for the background geometry. Furthermore, setting $\hSpBS$ as the background metric removes the need to specify a somewhat unphysical external metric. Indeed, the conformal mean metric $\hSpBS$ is determined by the dynamics itself and using it is a very natural choice.

\section{The dynamics of the geometric mean}
\label{sec:geometricmean}

In this section, we establish the evolution equation for the spatial part $\hSp$ of the geometric mean metric $\hMet$, and the relation between the extrinsic curvatures of the spatial metrics $\gSp,\fSp,\hSp$. These results are needed to compute the mean gauges in \autoref{sec:meangauges}.

\subsection{The evolution equation for $\chi$}
\label{subsec:EEchi}

In this subsection, we compute the evolution equation for the spatial part $\hSp$ of the geometric mean $\hMet$. The geometric mean does not satisfy the same field equations as $\gMet$ and $\fMet$ and it is not a dynamical variable. Hence, we need to express its derivative in terms of those of the dynamical variables.

The spatial part of the geometric mean is equal to \cite{MikicaB.794295},
\begin{align}
\label{eq:chi}
	\hSp = \gE^\tr\sEta\sLs\sRs\fEtr,
\end{align}
where $\gE,\fEtr$ are two freely specifiable vielbeins of the spatial parts $\gSp,\fSp$ of $\gMet,\fMet$ and only contain the metric fields, $\sLs$ is the spatial part of a Lorentz boost, and $\sRs$ is a Euclidean spatial rotation. The two transformations $\sLs$ and $\sRs$ in the Lorentz frame are not freely specifiable, since they are determined by:
\begin{enumerate}
	\item The requirement $\hSp=\hSp^\tr$ which, together with \eqref{eq:gaugevariablesshifts}, is equivalent to the requirement that a \emph{real} square root $\mSqrt\coloneqq\rb{\gMet^{-1}\fMet}^{1/2}$ exists \cite{MikicaB.794295}. This fixes $\sRs$ in terms of $\sLs$ (see Appendix~\ref{app:dynamicsdec}); note that both of $\sLs$ and $\sRs$ contain three degrees of freedom each, and the requirement $\hSp=\hSp^\tr$ fixes the degrees of freedom of $\sRs$ in terms of those of $\sLs$, contained in the spatial Lorentz vector $\sLp$, the separation parameter in \eqref{eq:gaugevariablesshifts}.
	\item The equations of motion. In \cite{HassanLundkvist2018,Kocic:2018ddp,Torsello_2019}, the three degrees of freedom in $\sLp$ are determined by solving one of the momentum constraints of the two metrics for $\sLp$. If we knew $\dt\sLp$ in terms of the dynamical variables, we could evolve $\sLp$ numerically without the need to solve the chosen momentum constraint at each time step. At present, we know $\sLp$ in spherical symmetry \cite{Kocic:2018ddp,Torsello_2019,Kocic_2019} and in the most general $\beta_{(1)}$-model \cite{Hassan:2011tf,HassanLundkvist2018}.
\end{enumerate}

Consider the differential operator $\pfh\coloneqq\dt-\mathcal{L}_\hShift = \mathcal{L}_{\hLapse\hnor}$, where $\hnor=\hLapse^{-1}(1,-\hShift)$ is the spacetime vector normal to the spacelike hypersurface with respect to $\hSp$, and $\hLapse\hnor$ is the normal evolution vector field \cite{gourgoulhon20123+1} with respect to $\hMet$. Applying $\pfh$ to $\hSp$ in \eqref{eq:chi} gives,
\begin{align}
\label{eq:EEchi}
	 \pfh\hSp	=\pfh \Bigrb{\gE^\tr \sEta \sLs \fE} 	&= \Bigrb{\pfh \gE^\tr} \sEta \sLs\sRs \fE + \gE^\tr \sEta \Bigrb{\pfh \sLs}\sRs \fEtr \nonumber \\
	 																	&\quad + \gE^\tr \sEta \sLs \Bigrb{\pfh \sRs} \fE + \gE^\tr \sEta \sLs\sRs \Bigrb{\pfh\fEtr}.
\end{align}
The time derivatives of the tetrads $\dt\gE^\tr$ and $\dt\fEtr$ are determined by the equation \cite{wald2010general,choquet2015introduction},
\begin{align}
\label{eq:eqtetrad}
	\partial_{[\mu}\bigl(E^\textbf{A}\bigr){}_{\nu]}=\bigl(E^\textbf{B}\bigr){}_{[\mu}\bigl(E^\textbf{C}\bigr){}_{\nu]}\spincon^\textbf{A}{}_{\textbf{BC}},
\end{align}
where $\textbf{A},\textbf{B},\textbf{C}$ are Lorentz indices running from 0 to 3, $\mu,\nu$ are spacetime indices, the parentheses $[\,]$ denote antisymmetrization of the indices they enclose, $\bigl(E^\textbf{A}\bigr){}_{\nu}$ is a 4-dimensional tetrad, and $\spincon^\textbf{A}{}_{\textbf{BC}}$ are the Ricci rotation coefficients, or connection coefficients, of the tetrad. To get the time derivative of the spatial part of $\bigl(E^\textbf{A}\bigr){}_{\nu}$, it is enough to set $\mu =0$ in \eqref{eq:eqtetrad}, which implies that $\nu$ is spatial, and $\textbf{A} = \textbf{a}$, with $\textbf{a}$ Lorentz index running from 1 to 3. This results in (see Appendix~\ref{app:dttetrads} for details),
\begin{align}
\label{eq:dertetrad}
	\bigl(\dt \gE^\textbf{a}\bigr){}_i= \dt\bigl(\gE^\textbf{a}{}_i\bigr),
\end{align}
i.e., we need to consider only the time derivative of the components of the \change{matrices of the tetrads}. The latter are given in terms of the metric functions of $\gSp$ and $\fSp$, and therefore the time derivative of the tetrads can be computed using the evolution equation for the spatial metrics. In addition, the Lie derivative of a tetrad is the Lie derivative of a covector \cite{Chinea_1988},
\begin{align}
	\Bigrb{\mathscr{L}_\hShift \gE^\textbf{a}}{}_{i} = \Bigrb{\gCD_i\hShift ^j}\gE^\textbf{a}{}_\nu+\hShift^j\Bigrb{\gCD_j\gE^\textbf{a}{}_i}, \quad \Bigrb{\mathscr{L}_\hShift \fE^\textbf{a}}{}_{i} = \Bigrb{\fCD_i\hShift ^j}\fE^\textbf{a}{}_\nu+\hShift^j\Bigrb{\fCD_j\fE^\textbf{a}{}_i}.
\end{align}
As shown in Appendix~\ref{app:tetrads}, this formula implies that the Lie derivative of a Lorentz tensor along a spacetime vector $X$ is equal to the directional derivative of its components along $X$. Hence, the Lie derivative sees Lorentz tensors as sets of scalars, and this is used below.

The time derivative of $\sRs$ can be expressed in terms of the derivatives of $\sLs$ and $\gE,\fEtr$, but the computation is quite technical and is relegated to Appendix~\ref{app:dynamicsdec}. Next, we consider the term $\pfh \sLs$ in \eqref{eq:EEchi}, which can be rewritten in terms of $\pfh \sLp$ by computing $\pfh\sLt$ first,
\begin{align}
\label{eq:dtsLt}
	\sLp' =\sLp^\tr\sEta, \quad \pfh\sLt		&=\pfh\sqrt{1+\sLp'\sLp}=\dfrac{1}{2}\dfrac{1}{\sqrt{1+\sLp'\sLp}}\qb{\brb{\pfh\sLp'}\sLp+\sLp'\brb{\pfh\sLp}}=\dfrac{1}{\sLt }\sLp'\pfh\sLp,
\end{align}
and plugging it into $\pfh \sLs$, which results in (see Appendix~\ref{app:dynamicsdec}),
\begin{align}
	\pfh\sLs		&=-\dfrac{1}{\sLt+1}\qb{\dfrac{\sLp'\pfh\sLp}{\rb{\sLt+1}\sLt}\sLp\sLp'-\bigrb{\pfh\sLp}\sLp'-\sLp\bigrb{\pfh\sLp'}}.
\end{align}
The problem reduces to the computation of $\pfh\sLp$, which is,
\begin{align}
\label{eq:pfp}
	\pfh\sLp		&= \dt \sLp - \hShift^i\partial_i\sLp.
\end{align}
As already discussed in \cite{Kocic:2018ddp,Torsello_2019,Kocic_2019}, the time derivative of the separation parameter $\sLp$ is not known explicitly in the general case. We do know it in the spherically symmetric case, though, hence in this case we can compute the time derivatives of $\sLs,\sRs$ and $\hSp$. The equations in spherical symmetry are presented in Appendix~\ref{app:meansgsph}, and were computed using the Mathematica package $\mathtt{bimEX}$ \cite{TORSELLO2020106948}.

We have computed the derivative of $\hSp$ along its normal evolution vector field $\hLapse\hnor=\mathrm{t}-\hShift$. This is the evolution equation necessary to compute the dynamics of the geometric mean sector and the mean gauges.

\subsection{The relation between the extrinsic curvatures}
\label{subsec:meanK}

In order to be able to compute the mean gauges, we also need an explicit expression for $\hK$, the trace of the extrinsic curvature of $\hSp$. This is computed by noting that the determinant of $\hSp$ is,\footnote{The determinant of $\sRs$ is just 1. Note that $\sRs$ is a proper rotation due to our choice of the principal branch of $(\gMet^{-1}\fMet)^{1/2}$.}
\begin{align}
\label{eq:dets}
	\hdet = \sLt\sqrt{\gdet\fdet}.
\end{align}
Let's now apply any derivative operator $\partial$ to \eqref{eq:dets}.\footnote{By \qm{any} derivative operator, we mean that the computation makes use of the Leibniz rule only.} We get (see Appendix \ref{app:cBSSNobservers}),
\begin{align}
\label{eq:observers}
	\dfrac{\partial \gdet}{\gdet} + \dfrac{\partial \fdet}{\fdet} = 2\rb{ \dfrac{\partial \hdet}{\hdet}-\dfrac{\sLp_\textbf{a}\partial \sLp^\textbf{a}}{\sLt^2} }.
\end{align}
If we set $\partial = \pfh$, the expression becomes,
\begin{align}
\label{eq:observerspf}
	\dfrac{\pfh \gdet}{\gdet} + \dfrac{\pfh \fdet}{\fdet} = 2\rb{ \dfrac{\pfh \hdet}{\hdet}-\dfrac{\sLp_\textbf{a}\pfh \sLp^\textbf{a}}{\sLt^2} }= 2\rb{-2\hLapse\hK-\dfrac{\sLp_\textbf{a}\pfh \sLp^\textbf{a}}{\sLt^2} },
\end{align}
where we used $\pfh\log(\hdet)=-2\hLapse\hK$ following from the definition of the extrinsic curvature.
Now consider the expressions on the left-hand side of \eqref{eq:observerspf}. Making use of \eqref{eq:gaugevariablesshifts}, the two terms therein can be rewritten as,
\begin{subequations}
\begin{align}
	\dfrac{\pfh \gdet}{\gdet} 	&=\dfrac{1}{\gdet}\rb{\dt \gdet-\mathcal{L}_\hShift \gdet}=\dfrac{1}{\gdet}\rb{\dt \gdet-\mathcal{L}_\gShift \gdet}+\dfrac{\mathcal{L}_{\gLapse \sLt^{-1}\sgn} \gdet}{\gdet}=-2\gLapse\gK+\dfrac{\mathcal{L}_{\gLapse \sLt^{-1}\sgn} \gdet}{\gdet}, \\
	\dfrac{\pfh \fdet}{\fdet}		&= -2\fLapse\fK-\dfrac{\mathcal{L}_{\fLapse \sLt^{-1}\sfn} \fdet}{\fdet}.
\end{align}
\end{subequations}
The Lie derivatives of the determinants, which are scalar densities of weight 2, are,
\begin{subequations}
\begin{align}
	\dfrac{\mathcal{L}_{\gLapse \sLt^{-1}\sgn} \gdet}{\gdet}	&=\gLapse \sLt^{-1}\sgn^i\partial_i\log\bigrb{\gdet}+2\,\partial_i\rb{\gLapse \sLt^{-1}\sgn^i}, \\
	\dfrac{\mathcal{L}_{\fLapse \sLt^{-1}\sfn} \fdet}{\fdet}		&=\fLapse \sLt^{-1}\sfn^i\partial_i\log\bigrb{\fdet}+2\,\partial_i\rb{\fLapse \sLt^{-1}\sfn^i}.
\end{align}
\end{subequations}
Hence, \eqref{eq:observerspf} becomes a relation between the traces of the extrinsic curvatures $\gK,\fK,\hK$,
\begin{align}
\label{eq:tracesK}
	\hK	&=\dfrac{1}{2\hLapse}\qb{\gLapse\gK-\dfrac{\mathcal{L}_{\gLapse \sLt^{-1}\sgn} \gdet}{2\gdet} +\fLapse\fK-\dfrac{\mathcal{L}_{\fLapse \sLt^{-1}\sfn} \fdet}{2\fdet}-\dfrac{\sLp_\textbf{a}\rb{\dt \sLp^\textbf{a}-\hShift^i\partial_i\sLp^\textbf{a}}}{\sLt^2}} \nonumber \\
			&=\dfrac{1}{2}\Biggl\{\sqrt{\sLt\mW}\Bigl[\gK- \dfrac{1}{2\sLt}\sgn^i\partial_i\log\bigl(\gdet\bigr)\Bigr]-\dfrac{1}{\hLapse}\partial_i\rb{\gLapse \sLt^{-1}\sgn^i}\nonumber \\
			&\quad\quad +\sqrt{\sLt\mW^{-1}}\qb{\fK+ \dfrac{1}{2\sLt}\sfn^i\partial_i\log\bigl(\fdet\bigr)}+\dfrac{1}{\hLapse}\partial_i\rb{\fLapse \sLt^{-1}\sfn^i}-\dfrac{\sLp_\textbf{a}\rb{\dt \sLp^\textbf{a}-\hShift^i\partial_i\sLp^\textbf{a}}}{\hLapse\sLt^2}\Biggr\}.
\end{align}
We want to isolate the mean lapse $\hLapse$ in \eqref{eq:tracesK}, so we use \eqref{eq:gaugevariableslapses} to replace the lapses $\gLapse$ and $\fLapse$ for $\hLapse$,
\begin{align}
\label{eq:exp}
	&\quad-\dfrac{1}{\hLapse}\partial_i\rb{\gLapse \sLt^{-1}\sgn^i-\fLapse \sLt^{-1}\sfn^i}=-\dfrac{1}{\hLapse}\partial_i\qb{\hLapse \rb{\sqrt{\mW\sLt^{-1}}\sgn^i-\sqrt{(\mW\sLt)^{-1}}\sfn^i}} \nonumber \\
	&=-\rb{\sqrt{\mW\sLt^{-1}}\sgn^i-\sqrt{(\mW\sLt)^{-1}}\sfn^i}\dfrac{\partial_i\hLapse}{\hLapse}-\partial_i\rb{\sqrt{\mW\sLt^{-1}}\sgn^i-\sqrt{(\mW\sLt)^{-1}}\sfn^i}.
\end{align}
The substitution of \eqref{eq:exp} into \eqref{eq:tracesK} yields,
\begin{align}
\label{eq:tracesKH}
	\hK	&=\dfrac{1}{2}\Biggl\{\sqrt{\sLt\mW}\Bigl[\gK- \dfrac{1}{2\sLt}\sgn^i\partial_i\log\rb{\gdet}\Bigr]+\sqrt{\sLt\mW^{-1}}\Bigl[\fK+ \dfrac{1}{2\sLt}\sfn^i\partial_i\log(\fdet)\Bigr]-\dfrac{\sLp_\textbf{a}\rb{\dt \sLp^\textbf{a}-\hShift^i\partial_i\sLp^\textbf{a}}}{\hLapse\sLt^2}\nonumber \\
			&\quad\quad -\rb{\sqrt{\mW\sLt^{-1}}\sgn^i-\sqrt{(\mW\sLt)^{-1}}\sfn^i}\dfrac{\partial_i\hLapse}{\hLapse}-\partial_i\rb{\sqrt{\mW\sLt^{-1}}\sgn^i-\sqrt{(\mW\sLt)^{-1}}\sfn^i}\Biggr\}.
\end{align}
In addition it holds,
\begin{align}
	\dfrac{\sLp_\textbf{a}\rb{\dt \sLp^\textbf{a}-\hShift^i\partial_i\sLp^\textbf{a}}}{\hLapse\sLt^2}=\mP^{i} \dfrac{\partial_i\hLapse}{\hLapse} +\mP_0,
\end{align}
since the evolution equation for $\sLp$, even though not known in general, must have a similar differential structure to (A.8) in \cite{Kocic:2018ddp}, i.e., to the evolution equation for the bimetric current $\gjotab$. Now we define the following functions of the dynamical fields not depending on $\hLapse$,
\begin{subequations}
\begin{align}
	\mF		&\coloneqq \sqrt{\sLt\mW}\Bigl[\gK - \dfrac{1}{2\sLt}\sgn^i\partial_i\log\rb{\gdet}\Bigr]+\sqrt{\sLt\mW^{-1}}\Bigl[\fK+ \dfrac{1}{2\sLt}\sfn^i\partial_i\log(\fdet)\Bigr] \nonumber \\
				&\quad-\partial_i\rb{\sqrt{\mW\sLt^{-1}}\sgn^i-\sqrt{(\mW\sLt)^{-1}}\sfn^i}-\mP_0, \\
	\mV^i	&\coloneqq -\sqrt{\mW\sLt^{-1}}\sgn^i+\sqrt{(\mW\sLt)^{-1}}\sfn^i-\mP^i,
\end{align}
\end{subequations}
such that,
\begin{align}
\label{eq:hK}
	\hK	&=\dfrac{1}{2}\Bigrb{\mV^i\partial_i\log(\hLapse)+\mF}.
\end{align}
The explicit expressions for $\mF$ and $\mV^i$ in spherical symmetry are given in Appendix~\ref{app:meansgsph}. Since \cite{baumgarte2010numerical},
\begin{align}
	\gK=-\gCD_\mu\gnor^\mu, \qquad \fK=-\fCD_\mu\fnor^\mu, \qquad \hK=-\hCD_\mu\hnor^\mu,
\end{align}
the equation \eqref{eq:tracesKH} is also a relation between the normal observers of $\gMet,\fMet,\hMet$, i.e., the observers having 4-velocity equal to the vectors $\gnor,\fnor,\hnor$ normal to the spacelike hypersurface with respect to $\gMet,\fMet,\hMet$.

\section{The mean gauges}
\label{sec:meangauges}

In this section we compute the standard gauge and the maximal slicing with respect to the geometric mean metric $\hMet=\gMet\bigrb{\gMet^{-1}\fMet}^{1/2}$. We choose to gauge fix the mean lapse function $\hLapse$ and the mean shift vector $\hShift$. The computations make use of the results in \autoref{sec:geometricmean}.

\subsection{The mean standard gauge}
\label{subsec:meansg}

The standard gauge in the cBSSN formulation of the EFE is \cite{PhysRevD.79.104029},
\begin{subequations}
\label{eq:standardgaugeGR}
	\begin{align}
		\label{eq:1plogGR}\partial_t \gLapse 		&= \gShift^j \gDback_j\gLapse -2\gLapse \gKBS, \\
		\label{eq:gdriver1GR}\partial_t \gShift^i	 	&= \gShift^j \gDback_j\gShift^i +\dfrac{3}{4}\gShiftB^i, \\
		\label{eq:gdriver2GR}\partial_t \gShiftB^i 	&= \gShift^j \gDback_j\gShiftB^i +\bigrb{\partial_t\gL^i -\gShift^i \gDback_j\gL^i} -\eta \gShiftB^i,
	\end{align}
\end{subequations}
consisting in the \onePlusLog gauge condition for the lapse \cite{PhysRevLett.75.600} and the $\Gamma$-driver gauge condition for the shift \cite{PhysRevD.67.084023}. In \eqref{eq:standardgaugeGR}, $\gShiftB$ is an auxiliary variable, $\gDback$ is the covariant derivative defined by the background connection, $\gL$ is the conformal connection, and $\eta$ is a free real constant.

We want to impose the standard gauge with respect to $\hMet$ and gauge fix $\hLapse,\hShift$. This means that we need to compute all the terms in the following equations,
\begin{subequations}
\label{eq:standardgauge}
	\begin{align}
		\label{eq:1plog}\partial_t \hLapse 		&= \hShift^j \hDback_j\hLapse -2\hLapse \hK, \\
		\label{eq:gdriver1}\partial_t \hShift^i	 	&= \hShift^j \hDback_j\hShift^i +\dfrac{3}{4}\hShiftB^i, \\
		\label{eq:gdriver2}\partial_t \hShiftB^i 	&= \hShift^j \hDback_j\hShiftB^i +\bigrb{\partial_t\hL^i -\hShift^i \hDback_j\hL^i} -\eta \hShiftB^i.
	\end{align}
\end{subequations}
In \eqref{eq:standardgauge}, $\hL^i\coloneqq \hSpBS^{jk} \hDG^i_{jk} = \hSpBS^{jk}\rb{\hGBS^i_{jk}-\hGbackBS^i_{jk}}$, where $\hL^i, \hGBS^i_{jk}$ are the conformal connection and the Christoffel symbol for $\hSpBS$, and $\hGbackBS^i_{jk}$ is the background connection for the mean sector, which is freely specifiable. We may choose one of the connections of $\gSpBS$ or $\fSpBS$, or a (local) affine combination of them,
\begin{align}
	\hGbackBS^i_{jk}= \sigma(x)\,\gGBS^i_{jk} + \bigqb{1-\sigma(x)} \,\fGBS^i_{jk}, \qquad 0\leq \sigma(x) \leq 1 \quad  \forall x,
\end{align}
as the background connection for the mean sector, since we know their evolution.\footnote{An affine combination is a linear combination whose coefficients sum up to 1. An affine combination of connections is a connection, contrary to a generic linear combination of them \cite[p.~52]{lee1997riemannian}.} Choosing $\sigma(x)=1/2$ gives the mean connection and is symmetric in $\gSpBS$ and $\fSpBS$. Following the computation outlined in \cite{PhysRevD.79.104029}, the evolution equation for the Christoffel symbol of the conformal metric $\gSpBS$ is found to be,
\begin{align}
\label{eq:dtGg}
	\dt\gGBS^i_{jk}	&=\mathscr{L}_\gShift\gGBS^i_{jk}-\dfrac{1}{6}\gSpBS^{ij}\partial_j\qb{\pfg\log(\gdetBS)-4\gLapse\gABS^i{}_i}+\gLapse\qb{\gABS^{jk}\gSpBS^{iq}\partial_q\gSpBS_{kj}-2\gABS^{kq}\gSpBS^{ij}\partial_q\gSpBS_{jk}},
\end{align}
and analogously for $\dt\fGBS^i_{jk}$ in the $\fMet$-sector. Hence, there is no need to choose an external connection in BR, since we have several metrics whose dynamics is known.

Now, we compute \eqref{eq:1plog} by substituting \eqref{eq:hK} in it,
\begin{align}
\label{eq:onePLusLogmean}
	\partial_t \hLapse 		&= \hShift^j \hDback_j\hLapse -2\hLapse \hK
		=\hShift^j \hDback_j\hLapse -\mV^i\partial_i\hLapse-\mF\hLapse.
\end{align}
Alternatively, we could also use the definition of $\hK_{ij}$ to get,
\begin{align}
\label{eq:hKdtchi}
	\hK=\hSp^{ij}\hK_{ij}= -\dfrac{1}{2}\hSp^{ij}\mathscr{L}_\hnor\hSp_{ij}=-\dfrac{1}{2\hLapse}\hSp^{ij}\pfh\hSp_{ij},
\end{align}
and then use the evolution equation \eqref{eq:EEchi} for $\hSp$. In this case, one should isolate the terms involving $\hLapse$ and its spatial derivatives in the evolution equation for $\hSp$. At this point, we have computed the \onePlusLog gauge condition with respect to $\hMet$, which we call \qm{mean \onePlusLog} gauge condition.

We now turn our attention to the \qm{mean $\Gamma$-driver} gauge condition. We need to compute $\dt\hL^i$ in terms of what we know, i.e., the evolution equation for the conformal mean metric $\hSpBS$. The derivative of $\hL^i$ along the normal evolution vector of $\hMet$ reads,
\begin{align}
\label{eq:dtL}
	\pfh\hL^i 	&=\dt\hL^i-\mathscr{L}_\hShift \hL^i=\pfh\Bigrb{ \hSpBS^{jk}\hDG^i_{jk}}=\pfh\Bigqb{\hSpBS^{jk}\Bigrb{\hGBS^i_{jk}-\hGbackBS^i_{jk}}},
\end{align}
implying,
\begin{align}
\label{eq:dtLfin}
	\dt\hL^i		&=\mathscr{L}_\hShift\hL^i+\hDG^i_{jk}\pfh\hSpBS^{jk}+\hSpBS^{jk}\pfh\hDG^i_{jk} \nonumber \\
					&=\mathscr{L}_\hShift\hL^i+\hDG^i_{jk}\pfh\hSpBS^{jk}+\hSpBS^{jk}\pfh\hGBS^i_{jk}-\hSpBS^{jk}\pfh\hGbackBS^i_{jk}.
\end{align}
In \eqref{eq:dtLfin}, the derivative of the inverse mean conformal metric is $\pfh\hSpBS^{ij}=-\hSpBS^{ik}\hSpBS^{j\ell}\pfh\hSpBS_{k\ell}$, and $\pfh\hSpBS_{k\ell}$ is the rewriting of $\eqref{eq:EEchi}$ in the cBSSN formalism, according to the conformal decomposition established in \cite{Torsello_2019}. The derivative of the Christoffel symbol of $\hSpBS$ is given by,
\begin{align}
\label{eq:dtG}
		\hSpBS^{jk}\pfh\hGBS^i_{jk}	&=-\hDBS_j\pfh\hSpBS^{ij}-\dfrac{1}{2}\hSpBS^{ij}\hDBS_j\pfh\log\bigrb{\hdetBS},			
\end{align}
following from the formula for the Lie derivative of the Christoffel symbol \cite[p.~291]{Blau:2017}. The latter also allows us how to deal with the derivative of the background connection,
\begin{align}
\label{eq:dtLback}
	\hSpBS^{jk}\pfh\hGbackBS^i_{jk}	&=\hSpBS^{jk}\dt\hGbackBS^i_{jk}-\hSpBS^{jk}\rb{\hDback_j\hDback_k\hShift^i-\hShift^\ell\hRback^i{}_{jk\ell}}.
\end{align}
In \eqref{eq:dtLback}, the time derivative is arbitrarily chosen. At this point, we know everything in \eqref{eq:standardgauge} and we can impose the standard gauge with respect to $\hMet$.

\subsection{The mean maximal slicing}
\label{subsec:meanms}

In this section we compute the gauge condition to impose the maximal slicing with respect to $\hMet$ in the standard \threePlusOne formalism,
\begin{equation}
\label{eq:meanmaxsli}
	\hK=\dt\hK=0.
\end{equation}
We use the expression for $\hK$ in \eqref{eq:hK} to impose $\hK=0$ on the initial hypersurface,
\begin{align}
\label{eq:meanmaxK}
	\mV^i\partial_i\log(\hLapse)+\mF=0.
\end{align}
This is already a boundary value problem for $\hLapse$, contrary to the analog case in GR. We can always choose the initial data to satisfy \eqref{eq:meanmaxK}. We take the time derivative of \eqref{eq:hK},
\begin{align}
\label{eq:meanmaxdtK}
	-2\,\dt\hK	&=\rb{\dt\mV^i}\partial_i\log(\hLapse)+\dt\mF+\mV^i\dfrac{\dt\partial_i\hLapse}{\hLapse}-\mV^i\partial_i\log(\hLapse)\,\dt\log(\hLapse),
\end{align}
and impose $\dt\hK=0$. We use \eqref{eq:meanmaxK} to rewrite the last term in \eqref{eq:meanmaxdtK},
\begin{align}
\label{eq:meanmaxeq}
	\mF\,\dt\log(\hLapse)+\rb{\dt\mV^i}\partial_i\log(\hLapse)+\dt\mF+\mV^i\dfrac{\dt\partial_i\hLapse}{\hLapse} =0.
\end{align}
This is the gauge condition for the mean maximal slicing that determines the lapse of $\hMet$. Note that, contrary to GR, this is not a boundary value problem. In addition, it is not a pure evolution equation in $\hLapse$, due to the presence of the term $\mV^i\dt\partial_i\hLapse$ with mixed time-spatial derivatives. Therefore, we need to rewrite \eqref{eq:meanmaxeq} in a way suitable for the numerical integration. It holds,
\begin{align}
\label{eq:Hders}
	\dfrac{\dt\partial_i\hLapse}{\hLapse}=\dfrac{\partial_i\dt\hLapse}{\hLapse}=\partial_i\dt\log(\hLapse)+\dt\log(\hLapse)\partial_i\log(\hLapse).
\end{align}
We insert \eqref{eq:Hders} in \eqref{eq:meanmaxeq} to get,
\begin{align}
\label{eq:meanmaxeq2}
	&\mF\,\dt\log(\hLapse)+\rb{\dt\mV^i}\partial_i\log(\hLapse)+\dt\mF \nonumber \\
	&+\mV^i\partial_i\dt\log(\hLapse)+\dt\log(\hLapse)\,\mV^i\partial_i\log(\hLapse) =0.
\end{align}
At this point, we define the two auxiliary variables,
\begin{align}
\label{eq:newvars}
	\heta	&\coloneqq \log(\hLapse), \qquad \hh \coloneqq \dt\heta,
\end{align}
and rewrite \eqref{eq:meanmaxeq2} as,
\begin{subequations}
\label{eq:meanmaxfinal}
	\begin{align}
		\dt\heta		&=\hh,\label{eq:meanmaxfinal1} \\
		0				&=\mV^i\partial_i\hh+\hh\rb{\mV^i\partial_i\heta+\mF}+\rb{\dt\mV^i}\partial_i\heta+\dt\mF.\label{eq:meanmaxfinal2}
	\end{align}
\end{subequations}
This is an evolution equation for $\heta$, constrained by a boundary value problem for its time derivative $\hh$. To solve it, first one solves \eqref{eq:meanmaxK} to get the initial value for $\heta$ such that $\hK=0$. Then, one determines $\hh$ on the initial hypersurface from \eqref{eq:meanmaxfinal2}, and finally evolves $\heta$ with \eqref{eq:meanmaxfinal1}. Note that \eqref{eq:meanmaxfinal2} contains the time derivatives of the dynamical fields, of $\sgn^i,\sfn^i$ and $\sLp^\textbf{a}$, and therefore contains first and second spatial derivatives of $\hLapse$, but not its time derivatives.

We stress that the mean maximal slicing needs to be studied in more detail, since it is the first gauge condition in BR which can hopefully be singularity avoiding for both metrics. Indeed, since the null-cones of $\hMet$ lie inside the convex hull of the null-cones of $\gMet$ and $\fMet$ \cite{HassanKocic2018}, the slicing made with respect to $\hMet$ is always \qm{in the middle} of those made with respect to $\gMet$ and $\fMet$. Moreover, it is not clear how to impose the maximal slicing simultaneously with respect to $\gMet$ and $\fMet$. This would lead us to impose simultaneously,
\begin{subequations}
\label{eq:bimmax1}
\begin{align}
	\gSp^{ij}\gD_i\gD_j\gLapse 	&= \gLapse \qb{\gK_{ij}\gK^{ij}+\dfrac{\gKappa}{2}\bigrb{\grhoeff+\gJotaeff^i{}_i}},\label{eq:gmax} \\
	\fSp^{ij}\fD_i\fD_j\fLapse 		&= \fLapse \qb{\fK_{ij}\fK^{ij}+\dfrac{\fKappa}{2}\bigrb{\frhoeff+\fJotaeff^i{}_i}},\label{eq:fmax}
\end{align}
\end{subequations}
each one being the usual maximal slicing condition in GR, one per metric sector. However, in BR the two lapses are not independent, $\gLapse=\mW\fLapse$, hence one has to substitute one lapse for the other in one of the equations in \eqref{eq:bimmax1}. As a consequence, we get a system of two partial differential equations for one unknown function. The study of the consistency of this system is left for future work. One possible way to start studying the system may be the use of the Cartan--Kuranishi prolongation algorithm \cite{10.2307/2372381} (see also \cite{seiler2009involution,jukka}), to complete the system \eqref{eq:bimmax1} to involutive form.

\section{The geometric mean in the covariant BSSN formalism}
\label{sec:meanBSSN}

In the cBSSN formulation of the EFE, one has the freedom to freely specify two things: the evolution of the determinant of the conformal metric and the background connection. In this section, we discuss the possibility to make these choices using the geometric mean metric $\hMet$ in the bimetric cBSSN formulation. We stress that these choices may affect the hyperbolicity of the system in BR. Since the bimetric cBSSN together with the standard gauge with respect to $\gMet$ or $\fMet$ is not proven to be well-posed, the possible choices described in this section represent opportunities to manipulate the hyperbolic structure to achieve well-posedness.

\subsection{The conservation laws for the determinants}
\label{subsec:observers}

In the cBSSN formulation of the EFE, the evolution of the determinant $\gdetBS$ of the conformal metric and the trace of the conformal extrinsic curvature $\gABS$ are not fixed but freely specifiable. In this subsection, we focus on the evolution of the determinant. Following \cite{PhysRevD.71.104011,Brown_2008,PhysRevD.79.104029}, we call \qm{Lagrangian} the case where the determinant is constant for an observer having 4-velocity $u^\mu=\mathrm{t}^\mu=(1,0,0,0)$, and \qm{Eulerian} the case where it is constant for an observer having 4-velocity $u^\mu=\gLapse\,\gnor^\mu=\bigrb{1,-\gShift^i}$, where $\gnor$ is the vector normal to the spacelike hypersurface with respect to $\gMet$. In formulas,
\begin{align}
	\dt \gdet =0, \qquad \pfg \gdet =\dt \gdet-\mathscr{L}_\gShift \gdet = \dt \gdet -2\gdet\gDBS_i\gShift^i =0.
\end{align}

In BR, since we have two metrics, we need to specify two conservation laws for the determinants. Here we show that there is a relation between the conservation laws associated to the conformal metrics $\gSpBS$, $\fSpBS$ and $\hSpBS$. An analog relation to \eqref{eq:dets} also holds for the determinant of the conformal metrics in the cBSSN formalism (see Appendix~\ref{app:cBSSNobservers}),
\begin{align}
\label{eq:BSSNdets}
	\hdetBS = \sLt\sqrt{\gdetBS\fdetBS},
\end{align}
to which we apply a derivative operator $\partial$ and get,
\begin{align}
\label{eq:BSSNdetsder}
	\dfrac{\partial \gdetBS}{\gdetBS} + \dfrac{\partial \fdetBS}{\fdetBS} = 2\rb{ \dfrac{\partial \hdetBS}{\hdetBS}-\dfrac{\sLp_\textbf{a}\partial \sLp^\textbf{a}}{\sLt^2} }.
\end{align}
This tells us that, once we make the free choices about the evolution of two of the three determinants $\gdetBS,\fdetBS,\hdetBS$, the evolution of the third one is completely specified by the evolution equation of the separation parameter $\sLp^\textbf{a}$, which is determined by the dynamics.

For definiteness, let's consider time derivatives. As our first choice, for example, we can set $\dt \hdetBS=0$, which implies,
\begin{align}
\dfrac{\dt \gdetBS}{\gdetBS} + \dfrac{\dt \fdetBS}{\fdetBS} = -2 \dfrac{\sLp_\textbf{a}\dt \sLp^\textbf{a}}{\sLt^2}.
\end{align}
We can now use the second free choice that we have, to fix one more of these derivatives. We can choose $\dt \gdetBS=-\gdetBS\rb{\sLp_\textbf{a}\dt \sLp^\textbf{a}}\sLt^{-2}$, which implies,
\begin{align}
\dfrac{\dt \gdetBS}{\gdetBS} = \dfrac{\dt \fdetBS}{\fdetBS} = - \dfrac{\sLp_\textbf{a}\dt \sLp^\textbf{a}}{\sLt^2}.
\end{align}
Therefore we have a natural choice, symmetric in the $\gMet$ and $\fMet$ sectors, that is, choosing the same evolution equation for the determinants of the conformal metrics. This is determined by the time evolution of $\sLp$, which is another reason why it would be desirable to be able to compute explicitly $\sLp$ or $\dt \sLp$ in the most general case. We call this choice the \qm{symmetric mean Lagrangian conservation law}, since it is the Lagrangian conservation law for the determinant of the mean conformal metric $\hSpBS$, and it is symmetric in the $\gMet$ and $\fMet$ sectors.

We can also define the \qm{symmetric mean Eulerian conservation law} by imposing $\pfh \hdetBS =0$. This implies,
\begin{align}
\dfrac{\pfh \gdetBS}{\gdetBS} + \dfrac{\pfh \fdetBS}{\fdetBS} = -2 \dfrac{\sLp_\textbf{a}\pfh \sLp^\textbf{a}}{\sLt^2},
\end{align}
which can be rewritten as,
\begin{align}
\biggrb{\dt \gdetBS -2\partial_i\hShift^i-\dfrac{\hShift^i\partial_i\gdetBS}{\gdetBS}} + 
\biggrb{\dt \fdetBS -2\partial_i\hShift^i-\dfrac{\hShift^i\partial_i\fdetBS}{\fdetBS}}  = -2 \dfrac{\sLp_\textbf{a}\pfh \sLp^\textbf{a}}{\sLt^2}.
\end{align}
We can make our second choice by setting the evolution of $\gdetBS$ to be,
\begin{align}
\dt \gdetBS 	&= 2\partial_i\hShift^i + \dfrac{1}{2}\hShift^i\rb{\dfrac{\partial_i\gdetBS}{\gdetBS}+\dfrac{\partial_i\fdetBS}{\fdetBS}}- \dfrac{\sLp_\textbf{a}\pfh \sLp^\textbf{a}}{\sLt^2},
\end{align}
which implies again $\dt \fdetBS = \dt \gdetBS$. The derivative $\pfh\sLp^\textbf{a}$ is given in \eqref{eq:pfp}.

Note that $\dt\sLp$ in spherical symmetry contains the spatial derivatives of the dynamical fields. Since, in the most general case, $\dt\sLp$ must have a similar differential structure to (A.8) in \cite{Kocic:2018ddp}, it should always contain the spatial derivatives of the dynamical fields. Hence, different evolution equations for the determinants $\gdetBS$ and $\fdetBS$ may affect the hyperbolicity of the system. The presence of the geometric mean sector provides more natural choices for the evolution equations of the determinants with respect to GR, and these choices can be seen as opportunities to recast the system in a well-posed form.

\subsection{The geometric mean as the background metric}
\label{subsec:meanback}

The evolution equation for the conformal connection in the cBSSN formalism is \cite{PhysRevD.79.104029,Torsello_2019},
\begin{align}
\label{eq:dtLg}
	\pfg \gL^i				&= -\gSpBS^{jk}\Bigrb{\dt\gGbackBS^i_{jk}-\mathscr{L}_\gShift\gGbackBS^i_{jk}}\nonumber \\
									&\quad\hspace{0.35mm} -\dfrac{1}{3}\gL^i\pfg\log\rb{\gdet}-\dfrac{1}{6}\gSpBS^{ij}\partial_j\pfg\log\rb{\gdet}-\dfrac{4}{3}\gLapse \gSp^{ij}\partial_j\gKBS \nonumber \\
									&\quad\hspace{0.35mm} -2 \Bigrb{\gABS^{jk}-\dfrac{1}{3}\gSpBS^{jk}\gABS}\Bigrb{\delta^i{}_j\partial_k\gLapse-6\gLapse\delta^i{}_j\partial_k\gconf-\gLapse \gDG^i_{jk}}-2\gKappa\gLapse\ee^{4\gconf}\gjotaeff^i,
\end{align}
with,
\begin{align}
	\mathscr{L}_\gShift\gGbackBS^i_{jk}	&=\gDback_j\gDback_k\gShift^i-\gRback^{i}{}_{jk\ell}\gShift^\ell.
\end{align}
Since the background connection is arbitrary, the term $\dt\gGbackBS^i_{jk}$ is set to zero in \cite{PhysRevD.79.104029}. In BR, though, we can set the background connections to $\hGBS^i_{jk}$, whose derivative along the normal evolution vector of $\gMet$ is,
\begin{align}
\label{eq:hback}
	\gSpBS^{jk}\rb{\dt\hGBS^i_{jk}-\mathscr{L}_\gShift\hGBS^i_{jk}}	&=\gSpBS^{jk}\rb{\dt\hGBS^i_{jk}-\mathscr{L}_\hShift\hGBS^i_{jk}+\mathscr{L}_\hShift\hGBS^i_{jk}-\mathscr{L}_\gShift\hGBS^i_{jk}} \nonumber \\
			&=\gSpBS^{jk}\rb{\pfh\hGBS^i_{jk}+\mathscr{L}_{\hShift-\gShift}\hGBS^i_{jk}}\nonumber \\
			&=\gSpBS^{jk}\qb{\pfh\hGBS^i_{jk}+\hDBS_j\hDBS_k\bigl(\hShift^i-\gShift^i\bigr)-\hRiBS^{i}{}_{jk\ell}\bigl(\hShift^\ell-\gShift^\ell\bigr)}\nonumber \\
			&=\sgBBS^j{}_{\ell}\,\hSpBS^{\ell k}\qb{\pfh\hGBS^i_{jk}-\hDBS_j\hDBS_k\bigl(\gLapse\sLt^{-1}\gE^i{}_\textbf{a}\sLp^\textbf{a})+\gLapse\sLt^{-1}\hRiBS^{i}{}_{jk\ell}\gE^\ell{}_\textbf{a}\sLp^\textbf{a}},
\end{align}
where we used $\gSpBS^{jk}=\sgBBS^j{}_{\ell}\,\hSpBS^{\ell k}$, with $\sgBBS^j{}_{\ell}=\ee^{2(\gconf-\fconf)}\sgB^j{}_{\ell}$, $\sgB^j{}_{\ell}=\gE^{-1}\sLs\sRs\fEtr$ (see \cite{Kocic:2018ddp,Torsello_2019}). It is,
\begin{align}
	\hSpBS^{\ell k}\pfh\hGBS^i_{jk}	&=-\dfrac{1}{2}\qb{\hDBS_j\pfh\hSpBS^{i\ell}+\hSpBS^{[i|m}\hDBS_m\bigrb{\hSpBS^{|\ell]k}\pfh\hSpBS_{jk}}},
\end{align}
again following from the formula for the Lie derivative of the Christoffel symbol. Hence, everything is known in \eqref{eq:hback} and the evolution equation \eqref{eq:dtLg} with $\hGBS^i_{jk}$ as the background connection can be integrated in time. An analog computation holds in the $\fMet$-sector.

Choosing $\hSpBS$ as the background metric for $\gSpBS$ and $\fSpBS$, removes the necessity to introduce an arbitrary external metric to make the conformal connection a vector under spatial coordinate transformation not involving the time coordinate. We remark that this also affects the hyperbolicity of the system, because the evolution equation for $\hSpBS$ is given in terms of the evolution equations of $\gSpBS,\fSpBS$ and $\sLp$.

There exists the possibility to choose simultaneously $\hSpBS$ as the background metric for $\gSpBS$ and $\fSpBS$, and the mean standard gauge. In that case, one of the connections of $\gSpBS$ or $\fSpBS$, or an affine combination of them, can be chosen as the background connection for $\hSpBS$. The evolution equations for the connections of $\gSpBS$ and $\fSpBS$ are given by \eqref{eq:dtGg} and the analog equation for the $\fMet$-sector. In this case, the system would be completely determined in terms of dynamical quantities only, without the need to introduce external geometries. How much this can be beneficial when one moves to the numerical integration, is an open question and the subject of ongoing work.

\section{Conclusions and outlook}
\label{sec:conclusions}

The main result of this paper is the computation of two gauge choices in bimetric relativity, with respect to the geometric mean $\hMet\coloneqq \gMet \operatorname{\#} \fMet = \gMet\,(\gMet^{-1}\fMet)^{1/2}=\gMet\,\mSqrt$, which we call \qm{mean gauges.} We computed the standard gauge and the maximal slicing, which are widely used in numerical relativity, with respect to $\hMet$. A priori, it was not obvious that this could be done, since the geometric mean is not a dynamical variable and does not satisfy the bimetric field equations. Therefore, the first step in our work consisted in computing the evolution equation for the spatial part $\hSp$ of $\hMet$ and the expression for the trace $\hK$ of its extrinsic curvature in terms of the dynamical variables. This allowed us to compute the mean gauges and gauge fix the mean lapse function $\hLapse$ and the mean shift vector $\hShift$. Since the gauge variables are related by \eqref{eq:gaugevariables}, this fixes the lapses and shifts of $\gMet$ and $\fMet$ as well.

The possibility to impose a gauge condition on $\hMet$ is important for at least two main reasons. First, we do not yet have a well-posed formulation of the \threePlusOne bimetric field equations, hence exploring different gauges which affect the hyperbolic structure is fundamental; this is the motivation to establish the mean standard gauge, whose effect on the hyperbolic structure is left for future work. Second, we would like to find a gauge which is singularity avoiding for both metrics; this is the motivation to determine the mean maximal slicing, since it is not clear if one can impose the maximal slicing with respect to $\gMet$ and $\fMet$ simultaneously. Being the geometric mean of $\gMet$ and $\fMet$, the hope is that imposing a singularity avoiding gauge on $\hMet$ will also be singularity avoiding for $\gMet$ and $\fMet$.

The maximal slicing with respect to $\hMet$ results in a constrained evolution equation for the logarithm of the mean lapse function $\hLapse$. The equation is such that one cannot readily state if imposing the maximal slicing on $\hMet$, which is singularity avoiding for $\hMet$, is also singularity avoiding for $\gMet$ and $\fMet$, since the traces of the extrinsic curvatures of $\gMet$ and $\fMet$ are not constrained to be zero. This means that we cannot exclude that the observers normal with respect to $\gMet$ and $\fMet$, i.e., the observers having 4-velocities equal to the vectors normal to the spacelike hypersurface with respect to $\gMet$ and $\fMet$, will follow focusing geodesics when a singularity is forming.

The mean standard gauge yields evolution equations for the mean lapse function $\hLapse$ and the mean shift vector $\hShift$. In general relativity, the standard gauge is used together with the (covariant) BSSN formulation because the complete system is strongly hyperbolic \cite{PhysRevD.66.064002,PhysRevD.70.104004,PhysRevD.79.104029}. In bimetric relativity, we cannot yet state that choosing the standard gauge with respect to $\gMet$ or $\fMet$ implies strong hyperbolicity \cite{Torsello_2019}. The mean standard gauge affects the hyperbolicity of the system, and is therefore an opportunity to recast it in a well-posed form. Also, this gauge choice gives us the freedom to choose an arbitrary background connection for the geometric mean sector. This choice can be used to alter the hyperbolicity. The study of the various possibilities is left for future work.

In the last part of the paper, we showed how the geometric mean can be used to change the hyperbolic structure of the system by using it as the background metric for the conformal metrics $\gSpBS$ and $\fSpBS$ in the covariant BSSN formalism. This is the natural continuation of the work made in \cite{Torsello_2019}. On the same line, we also showed that the various conservation laws for the determinants of the conformal metrics that one has the freedom to impose in the covariant BSSN formalism, are related in bimetric relativity. If we choose the evolution equations for the determinants of two of the three conformal metrics $\gSpBS,\fSpBS,\hSpBS$, the evolution equation for the third one is determined by the evolution equation for the separation parameter $\sLp$. This equation also contains the spatial derivatives of the dynamical fields, and therefore the choice of the conservation laws in bimetric relativity is another opportunity to recast the system in a well-posed form.

The mean gauges and the choices regarding the conservation laws for the determinants of the metrics and the background connections in the bimetric covariant BSSN formalism need to be studied more in two respects:
\begin{enumerate}
	\item The study of the hyperbolicity for all the possible combinations of these choices (e.g., the mean standard gauge with the mean metric being the background metric for $\gSpBS$ and $\fSpBS$, and the mean Lagrangian conservation law) must be studied theoretically, as well as the singularity avoidance of the mean maximal slicing with respect to $\gMet,\fMet$.
	\item The numerical advantages of these choices have to be tested. This is connected to the previous point about well-posedness, but it has a separate value, since a different form of the equations can introduce different sources of numerical errors.
\end{enumerate}

\subsection*{Acknowledgments}

This work could not be completed without all the profitable discussions with Mikica Kocic, Edvard M\"ortsell, Anders Lundkvist and Marcus H\"ogås. I am pleased to thank Edvard M\"ortsell, Mikica Kocic and Marcus Högås for reading the paper and providing valuable comments and remarks.

\appendix

\section{Notation}
\label{app:notation}

In this paper we consider three metric sectors, the Lorentz frame and their BSSN formulation. The notation referring to each of these sectors is,
\begin{alignat}{4}
	\gG 		&\ ,\quad &\mbox{no accent} &:\quad &&\mbox{object refers to the $g$-sector},& \nonumber \\
	\fG			&\ ,\; &\mbox{tilde}&:&& \mbox{object refers to the $f$-sector},& \nonumber \\
	\hG 		&\ ,\; &\mbox{hash}&: &&\mbox{object refers to the $h$-sector},& \nonumber \\
	\lSector{\boldsymbol{\Gamma}} 		&\ ,\; &\mbox{boldface}&: &&\text{object refers to the Lorentz frame},& \nonumber \\
	\gGBS 		&\ ,\quad &\mbox{overbar} &:\quad &&\mbox{object refers to the $g$-sector in BSSN},& \nonumber \\
	\fGBS 			&\ ,\; &\mbox{wide hat}&:&& \mbox{object refers to the $f$-sector in BSSN},& \nonumber \\
	\hGBS 		&\ ,\; &\mbox{circle}&: &&\mbox{object refers to the $h$-sector in BSSN},& \nonumber \\
	\lSector{\lBSSN{\boldsymbol{\Gamma}}} 		&\ ,\; &\mbox{boldface, asterisk}&: &&\mbox{object refers to the Lorentz frame in BSSN}.& \nonumber 
\end{alignat}
Tensors are written both with and without indices, e.g., the metric $\fMet$ or $\fMet _{\mu\nu}$. Greek indices run from 0 to 3; latin indices run from 1 to 3; uppercase boldface indices are Lorentz indices from 0 to 3; lowercase boldface indices are spatial Lorentz indices from 1 to 3.

\section{Explicit computations and equations}
\label{app:appendix}

\subsection{The time derivative of the tetrads}
\label{app:dttetrads}

In this appendix we clarify how to obtain \eqref{eq:dertetrad} from \eqref{eq:eqtetrad}. To compute $\spincon^\textbf{A}{}_{\textbf{BC}}$ in \eqref{eq:eqtetrad}, we note that the Cauchy adapted frames and coframes \cite{choquet2015introduction} for $\gMet$ and $\fMet$ are \cite{Kocic:2018ddp},
\begin{subequations}
\label{eq:cauchy}
\begin{alignat}{6}
	\gcauchy^0 &= \dd t, & \qquad \gcauchy^i	&=\dd x^i+\gShift^i\dd t, &\qquad\qquad \gcauchy_0 &= \partial_t-\gShift^i\partial_i, & \qquad \gcauchy_i	&=\partial_i, \\
	\fcauchy^0 &= \dd t, 	& \qquad \fcauchy^i 	&=\dd x^i+\fShift^i\dd t, &\qquad\qquad \fcauchy_0 &= \partial_t-\fShift^i\partial_i, & \qquad \fcauchy_i	&=\partial_i,
\end{alignat}
\end{subequations}
and the 4-dimensional tetrads of $\gMet$ and $\fMet$ can be written as \cite{Kocic:2018ddp},
\begin{subequations}
\label{eq:tetrads}
\begin{alignat}{5}
	\gFE^\textbf{0} &=\gLapse\, \gcauchy^0			&&= \gLapse\,\dd t, & \qquad \gFE^\textbf{a} &= \gE^\textbf{a}{}_i\gcauchy^i &&= \gE^\textbf{a}{}_i \bigl( \dd x^i+\gShift^i\dd t \bigr), \\
	\gFE_\textbf{0} &=\gLapse^{-1} \gcauchy_0		&&= \gLapse^{-1}\bigl( \partial_t-\gShift^i\partial_i \bigr), & \qquad \gFE_\textbf{a} &= \gE^i{}_\textbf{a}\gcauchy_i &&= \gE^i{}_\textbf{a} \partial_i, \\
	\fFEtr^\textbf{0} &=\fLapse\, \fcauchy^0			&&= \fLapse\,\dd t, & \qquad \fFEtr^\textbf{a} &= \fEtr^\textbf{a}{}_i\fcauchy^i &&= \fEtr^\textbf{a}{}_i \bigl( \dd x^i+\fShift^i\dd t \bigr), \\
	\fFEtr_\textbf{0} &=\fLapse^{-1} \fcauchy_0		&&= \fLapse^{-1}\bigl( \partial_t-\fShift^i\partial_i \bigr), & \qquad \fFEtr_\textbf{a} &= \fEtr^i{}_\textbf{a}\fcauchy_i &&= \fEtr^i{}_\textbf{a} \partial_i.
\end{alignat}
\end{subequations}
We write \eqref{eq:tetrads} collectively as,
\begin{align}
	\gFE^\textbf{A} = \gEm^\textbf{A}{}_\mu \gcauchy^\mu, \qquad \gFE_\textbf{A} = \gEm^\mu{}_\textbf{A} \gcauchy_\mu, \qquad \fFEtr^\textbf{A} = \fEm^\textbf{A}{}_\mu \fcauchy^\mu, \qquad \fFEtr_\textbf{A} = \fEm^\mu{}_\textbf{A} \fcauchy_\mu,
\end{align}
such that the connection coefficients for the tetrads are,
\begin{align}
\label{eq:spinconexp}
	\gspincon^\textbf{A}{}_{\textbf{BC}}=\gEm^\textbf{A}{}_\sigma\,\gEm^\alpha{}_\textbf{B}\,\gEm^\rho{}_\textbf{C}\,\gcauchycon^\sigma{}_{\rho\alpha}+\bigl(\gFE^\textbf{A}\bigr){}_\sigma\bigl( \partial_\rho\gEm^\alpha{}_\textbf{B} \bigr)\bigl(\gFE_\textbf{C}\bigr)^\rho\bigl( \gcauchy_\alpha \bigr)^\sigma,
\end{align}
and analogously for the $\fMet$-sector. The connection coefficients for the Cauchy frame $\gcauchycon^\sigma{}_{\rho\alpha}$ can be found in \cite{choquet2015introduction}. Hence, by using \eqref{eq:spinconexp}, one can expand \eqref{eq:eqtetrad} to get \eqref{eq:dertetrad}. The computation in the $\fMet$-sector is analogous.

\subsection{The Lie derivatives of Lorentz tensors}
\label{app:tetrads}

Consider the tetrad---i.e., the Lorentz coframe---$(\gFE^\textbf{a}){}_\mu$ in which the metric $\gMet_{\mu\nu}$ can be written as
\begin{align}
\gMet_{\mu\nu}=(\gFE^\textbf{a})_\mu\Eta_{\textbf{a}\textbf{b}}(\gFE^\textbf{b})_\nu.
\end{align}
Its Lie derivative along a vector field $X^\mu$ is given by \cite{Chinea_1988},
\begin{align}
\label{eq:Lietetrad}
	\mathscr{L}_{X^\nu} (\gFE^\textbf{a}){}_\mu &=\qb{\gD_\mu X^\nu}(\gFE^\textbf{a})_\nu+X^\nu\qb{\gD_\nu (\gFE^\textbf{a})_\mu},
\end{align}
i.e., the Lie derivative of a covector. The Lorentz index is a label indicating which of the 1-forms in the Lorentz coframe we are considering. Therefore, the Lie derivative is blind to the index \textbf{a}.

The formula \eqref{eq:Lietetrad} allows us to compute the Lie derivatives of Lorentz tensors in a straightforward way. We consider only a Lorentz vector here, but the method can be generalized to Lorentz tensors of any rank. First, we note that the definition of the Lorentz coframe $(\gFE^\textbf{a})_\mu$ implies the existence of the dual Lorentz frame $(\theta_\textbf{a})^\mu$ such that
\begin{align}
	\theta_\textbf{a}\bigqb{\gFE^\textbf{b}}= (\theta_\textbf{a})^\mu(\gFE^\textbf{b}){}_\mu = \delta_\textbf{a}{}^\textbf{b}.
\end{align}
Second, consider the Lorentz vector $U^\textbf{a}$; we can map it to a spacetime vector by using the Lorentz frame and coframe
\begin{align}
\label{eq:dualvectors}
	V^\mu = (\theta_\textbf{a})^\mu U^\textbf{a},\qquad  U^\textbf{a} = (\gFE^\textbf{a})_\mu V^\mu.
\end{align}
Using \eqref{eq:dualvectors}, the Lie derivative of $U^\textbf{a}$ along the spacetime vector $X^\mu$ becomes,
\begin{align}
	\mathscr{L}_{X^\nu} U^\textbf{a}	&=\mathscr{L}_{X^\nu}\qb{(\gFE^\textbf{a})_\mu V^\mu} = \qb{\mathscr{L}_{X^\nu}(\gFE^\textbf{a})_\mu}V^\mu + (\gFE^\textbf{a})_\mu\qb{\mathscr{L}_{X^\nu}V^\mu} \nonumber \\
														&=\cb{\qb{\partial_\mu X^\nu}(\gFE^\textbf{a})_\nu+X^\nu\qb{\partial_\nu (\gFE^\textbf{a})_\mu}}V^\mu + (\gFE^\textbf{a})_\mu \cb{X^\nu\partial_\nu V^\mu-V^\nu\partial_\nu X^\mu}  \nonumber \\
														&=\cb{V^\mu\qb{\partial_\mu X^\nu}(\gFE^\textbf{a})_\nu-V^\nu\qb{\partial_\nu X^\mu}(\gFE^\textbf{a})_\mu}+X^\nu\qb{\partial_\nu (\gFE^\textbf{a})_\mu}V^\mu + (\gFE^\textbf{a})_\mu X^\nu\partial_\nu V^\mu   \nonumber \\
														&=X^\nu\cb{\qb{\partial_\nu (\gFE^\textbf{a})_\mu}V^\mu + (\gFE^\textbf{a})_\mu \partial_\nu V^\mu }=X^\nu\partial_\nu\qb{(\gFE^\textbf{a}{}_\mu V^\mu)}= X^\nu\partial_\nu U^\textbf{a},
\end{align}
which is the Lie derivative of a scalar function. Again, the Lie derivative is blind to the Lorentz index, and sees $U^\textbf{a}$ as a set of four scalar functions [and $(E^\textbf{a})_\mu$ as a set of four 1-forms]. This proves the formula in \eqref{eq:pfp}.

\subsection{The dynamics of the bimetric Lorentz frame}
\label{app:dynamicsdec}

In this \change{appendix}, we assume to know an expression for $\dtp$ and
express the time derivatives of $\sLs$ and $\sRs$ in terms of it. If the expression for $\dtp$
was known in general\textemdash it is known in spherical symmetry \cite{Kocic:2018ddp,Torsello_2019,Kocic_2019} and in the most general $\beta_{(1)}$-model \cite{Hassan:2011tf,HassanLundkvist2018}\textemdash one
could solve one momentum constraint for $\sLp$ on the spacelike hypersurface
where the initial data are specified, and then integrate the evolution
equation for $\sLp$. In addition, as shown in \autoref{subsec:observers}, $\dtp$
determines the relation between the conservation laws for the determinants of the conformal metrics in
the bimetric cBSSN formalism.

Following \cite{Kocic:2018ddp}, given an arbitrary $\sLp$ and a
freely specifiable spatial vielbein $\fEtr$ for the spatial part $\fSp$
of $\fMet$, we can find another vielbein $\fE$ of $\fSp$ such that
$\hSp=\gE^{\intercal}\sEta\sLs\fE=\hSp^{\intercal}$. The new vielbein
$\fE$ is determined in the following way, 
\begin{equation}
\fE\coloneqq\sRs\,\fEtr,\quad\sRs=\rb{\sEta^{-1}\sRbar^{\tr}\sEta\sRbar}^{1/2}\sRbar^{-1},\quad\sRbar\coloneqq\sEta^{-1}\fEtr^{-1,\tr}\gE^{\tr}\sEta\sLs,\label{eq:m}
\end{equation}
where $\sRs$ is the orthogonal transformation appearing in the polar
decomposition of $\sRbar^{-1}$,\footnote{Concerning the polar decomposition of a matrix, we refer to \cite[Sec.~2.5]{hall2015lie}.}
\begin{equation}
\sRbar^{-1}=\rb{\sRbar^{-1}\sEta^{-1}\sRbar^{-1,\tr}\sEta}^{1/2}\sRs=\rb{\sEta^{-1}\sRbar^\tr\sEta\sRbar}^{-1/2}\sRs.
\end{equation}
Therefore, we can express $\dt\sRs$ in terms of $\dt\sLs$ by using
\eqref{eq:m}, and then express $\dt\sLs$ in terms of $\dtp$ via
\begin{equation}
\sLs=\sI+\dfrac{1}{\sLt+1}\sLp\sLp^{\tr}\sEta,\quad\sLt=\rb{1+\sLp^{\tr}\sEta\sLp}^{1/2}.\label{eq:sLs}
\end{equation}
This provides the evolution equations for $\sRs$ and $\sLs$ in terms
of $\dtp$.

Let us start with relating $\dt\sLs$ and $\dtp$. A direct computation of the time derivative of \eqref{eq:sLs} gives, 
\begin{align}
\label{eq:dtsLs}
	\dt\sLs		&=\dt\rb{\sI+\dfrac{1}{\sLt+1}\sLp\sLp^\tr\sEta}=\dt\sI+\dt\rb{\dfrac{1}{\sLt+1}}\sLp\sLp'+\dfrac{1}{\sLt+1}\dt\rb{\sLp\sLp'} \nonumber \\
					&=-\dfrac{\dt\sLt}{\rb{\sLt+1}^2}\sLp\sLp'+\dfrac{1}{\sLt+1}\qb{\bigrb{\dt\sLp}\sLp'+\sLp\bigrb{\dt\sLp'}} \nonumber \\
					&=-\dfrac{1}{\sLt+1}\qb{\dfrac{\sLp'\dt\sLp}{\rb{\sLt+1}\sLt}\sLp\sLp'-\bigrb{\dt\sLp}\sLp'-\sLp\bigrb{\dt\sLp}'},
\end{align}
where we introduced the adjoint $\sLp'=\sLp^{\tr}\sEta$ of $\sLp$,
and used $\dt\sLp'=\bigrb{\dtp}'$ and \eqref{eq:dtsLt} with $\pfh\rightarrow \dt$. We now express $\dt\sRbar$ in terms of $\dt\sLs$, 
\begin{align}
\sRbar\coloneqq\sEta^{-1}\fEtr^{-1,\tr}\gE^{\tr}\sEta\sLs & \Longrightarrow\dt\sRbar=\dt\rb{\sEta^{-1}\fEtr^{-1,\tr}\gE^{\tr}\sEta}\sLs+\sEta^{-1}\fEtr^{-1,\tr}\gE^{\tr}\sEta\,\dt\sLs,%
\end{align}
where the time derivatives of $\gE$ and $\fEtr$ only contain the
time derivatives of the metric functions, and are then determined
by the evolution equations for the metrics. At this point, we need
to write $\dt\sRs$ as a function of $\dt\sRbar$. The two linear
operators are related by, 
\begin{align}
\sRs=\rb{\sRbar'\sRbar}^{1/2}\sRbar^{-1}\Longrightarrow\sRs\sRbar=\rb{\sRbar'\sRbar}^{1/2},
\end{align}
where $\sRbar'=\sEta^{-1}\sRbar^{\tr}\sEta$ is the adjoint of $\sRbar$.
We take the time derivative and obtain, 
\begin{align}
\dt\rb{\sRs\sRbar} & =\dt\rb{\sRbar'\sRbar}^{1/2} \Longrightarrow \rb{\dt\sRs}\sRbar+\sRs\,\dt\sRbar  =\dt\rb{\sRbar'\sRbar}^{1/2},
\end{align}
which implies, 
\begin{align}
\dt\sRs & =\qb{\dt\rb{\sRbar'\sRbar}^{1/2}}\sRbar^{-1}-\rb{\sRs\,\dt\sRbar}\sRbar^{-1}.\label{eq:dtsRs}
\end{align}
We now need to express $\dt\rb{\sRbar'\sRbar}^{1/2}$ in terms of
$\dt\sRbar$. It holds, 
\begin{align}
\label{eq:SylR}
\dt\rb{\sRbar'\sRbar}	&=\dt\qb{\rb{\sRbar'\sRbar}^{1/2}\rb{\sRbar'\sRbar}^{1/2}} \nonumber \\
								&=\rb{\sRbar'\sRbar}^{1/2}\dt\qb{\rb{\sRbar'\sRbar}^{1/2}}+\dt\qb{\rb{\sRbar'\sRbar}^{1/2}}\rb{\sRbar'\sRbar}^{1/2}.
\end{align}
This is a subcase of the Sylvester matrix equation 
\begin{align}
\label{eq:Syl}
AX+XB=C,
\end{align}
where $X$ is the unknown matrix.\footnote{The Sylvester equation also appears in BR when varying the square root $\mSqrt$, see Appendix B in \cite{Bernard:2015mkk}.} The Sylvester equation \eqref{eq:Syl} has a unique
solution for every $C$, if and only if $A$ and $-B$ have no common
eigenvalues \cite{ROTELLA1989271}. In our case $A=B=\rb{\sRbar'\sRbar}^{1/2}$,
$X=\dt\bigqb{\rb{\sRbar'\sRbar}^{1/2}}$, $C=\dt\rb{\sRbar'\sRbar}$.
Since $A=\rb{\sRbar'\sRbar}^{1/2}$ is strictly positive definite,\footnote{The polar decomposition of the invertible matrix $\sRbar^{-1}$ guarantees
that \cite[Sec.~2.5]{hall2015lie}.} its eigenvalues are strictly positive and $B=-\rb{\sRbar'\sRbar}^{1/2}$
has strictly negative eigenvalues. This implies that the equation \eqref{eq:SylR}
has a unique solution for $\dt\bigqb{\rb{\sRbar'\sRbar}^{1/2}}$. The general solution of the Sylvester equation
can be written very simply by making use of the vectorization operator.
For this reason, we now briefly introduce it and refer the reader
to \cite{magnus2007matrix} for a more extended treatment.

Vectorization $\vect{\cdot}$ is an isomorphism between $\mathbb{C}^{n\times m}$
and $\mathbb{C}^{nm}$. It corresponds to taking the columns of a
matrix and arranging them in order, within a column vector, e.g.,
\begin{align}
\begin{pmatrix}a & b\\
c & d
\end{pmatrix}\overset{\op{vec}}{\longrightarrow}\begin{pmatrix}a\\
c\\
b\\
d
\end{pmatrix}.
\end{align}
Some useful properties of this operator are \cite[Sec.~9]{MACEDO20132160},
\bSe \label{eq:vecprop} 
\begin{alignat}{4}
\vect{AB}	 	& =\rb{I\otimes A}\vect{B}\; 	&=\,	&\rb{B^{\tr}\otimes I}\vect A,\\
\vect{ABC} 	& =\rb{I\otimes AB}\vect{C}\;	&=\,	&\rb{C^{\tr}B^{\tr}\otimes I}\vect A=\rb{C^{\tr}\otimes A}\vect B,
\end{alignat}
\eSe
where $\otimes$ denotes the Kronecker product (see \cite[Sec. 4.2]{horn1994topics}). The Kronecker product of two matrices $A \in \mathbb{C}^{m\times n}$ and $B \in \mathbb{C}^{p\times q}$ is equal to,
\begin{align}
	A\otimes B = \mat{
		A_{11}\,B		&\cdots 		& A_{1n}\,B \\	
		\vdots 			&\ddots 	& \vdots \\
		A_{m1}\,B	 	&\cdots 		& A_{mn}\,B \\
	} \in \mathbb{C}^{mp\times nq}.
\end{align}
The general solution
of the Sylvester equation reads \cite{ROTELLA1989271}, 
\begin{align}
\vect X=\rb{I\otimes A+B^{\tr}\otimes I}^{-1}\vect C.
\end{align}
In our case it corresponds to,
\begin{equation}
\vect{\dt{\rb{\sRbar'\sRbar}^{1/2}}}=\cb{\sI\otimes\rb{\sRbar'\sRbar}^{1/2}+\qb{\rb{\sRbar'\sRbar}^{1/2}}^{\tr}\otimes\sI}^{-1}\vect{\dt\rb{\sRbar'\sRbar}}.\label{eq:Sylv}
\end{equation}
Lastly, we express $\vect{\dt\rb{\sRbar'\sRbar}}$ in terms of $\vect{\dt\sRbar}$.
Using the facts that the time derivative commutes with $\vect{\cdot}$,
that $\vect{A^{\tr}}=\mathsf{K}\vect A$ with $\mathsf{K}$ \qm{commutation matrix}
\cite{magnus2007matrix}, and using \eqref{eq:vecprop}, one gets,
\begin{align}
\vect{\dt\rb{\sRbar'\sRbar}} & =\qb{\rb{\sRbar^{\tr}\otimes\sI}\rb{\sEta\otimes\sEta^{-1}}\mathsf{K}+\rb{\sI\otimes\sRbar'}}\vect{\dt\sRbar}.
\end{align}
Hence, if we know $\dtp$, we can evolve all the objects in the bimetric
decomposition. Indeed, the latter depend on $\sRs,\sLs,\gE,\fEtr$,
whose dynamics depend on $\dtp$ and the evolution equations for the
metrics. Ultimately, since $\dtp$ itself depends on the evolution of the dynamical variables, the dynamics of all the objects in the bimetric decomposition depends only on the evolution of the dynamical variables.

\subsection{The relation between the determinants of the conformal metrics}
\label{app:cBSSNobservers}

In this appendix, we derive \eqref{eq:BSSNdets}, \eqref{eq:BSSNdetsder} and \eqref{eq:observers}. We start with \eqref{eq:dets},
\begin{align}
	\hdet = \sLt\sqrt{\gdet\fdet},
\end{align}
and rewrite it in the BSSN formalism,
\begin{align}
\label{eq:reldets}
	\ee^{6\rb{\gconf+\fconf}}\hdetBS = \sLt\sqrt{\ee^{12\gconf}\gdetBS\ee^{12\fconf}\fdetBS}\Longrightarrow \hdetBS = \sLt\sqrt{\gdetBS\fdetBS},
\end{align}
i.e., we get \eqref{eq:BSSNdets}. We now apply any derivative operator, denoted here by $\partial$, to \eqref{eq:reldets},
\begin{align}
	\partial\hdetBS = \partial\rb{\sLt\sqrt{\gdetBS\fdetBS}}=\dfrac{\sLt}{2}(\gdetBS\fdetBS)^{-1/2}\rb{\fdetBS\partial\gdetBS+\gdetBS\partial\fdetBS}+(\gdetBS\fdetBS)^{1/2}\partial \sLt.
\end{align}
Using \eqref{eq:dtsLt} with $\pfh$ replaced by $\partial$, we get,
\begin{align}
\label{eq:obstemp}
	\partial\hdetBS = \partial\rb{\sLt\sqrt{\gdetBS\fdetBS}}=\dfrac{\sLt}{2(\gdetBS\fdetBS)^{1/2}}\rb{\fdetBS\partial\gdetBS+\gdetBS\partial\fdetBS}+(\gdetBS\fdetBS)^{1/2}\dfrac{1}{\sLt}\sLp^\tr\sEta\partial\sLp.
\end{align}
Now divide equation \eqref{eq:obstemp} by $\hdetBS = \sLt(\gdetBS\fdetBS)^{1/2}$,
\begin{align}
	\dfrac{\partial\hdetBS}{\hdetBS} =\dfrac{1}{2\gdetBS\fdetBS}\rb{\fdetBS\partial\gdetBS+\gdetBS\partial\fdetBS}+\dfrac{1}{\sLt^2}\sLp^\tr\sEta\partial\sLp,
\end{align}
which can be rewritten as
\begin{align}
	\dfrac{\partial \gdetBS}{\gdetBS} + \dfrac{\partial \fdetBS}{\fdetBS} = 2\rb{ \dfrac{\partial \hdetBS}{\hdetBS}-\dfrac{\sLp^\tr\sEta\partial \sLp}{\sLt^2} },
\end{align}
i.e., the formula in \eqref{eq:BSSNdetsder}. The same steps lead from \eqref{eq:dets} to \eqref{eq:observers}.

\subsection{The mean gauges in spherical symmetry}
\label{app:meansgsph}

In this \change{appendix} we reduce to spherical symmetry the most relevant equations introduced in the main text. The computations were done with the Mathematica package $\mathtt{bimEX}$ \cite{TORSELLO2020106948}.

The spherically symmetric ansatz with the metric $\gMet$ and $\fMet$ sharing the same Killing vector fields is \cite{Torsello_2019},
\begin{alignat}{4}
	\gEBS^\textbf{a}{}_i 	&= \diagmat{\gA (t,r),\gB (t,r) ,\gB (t,r)\sin (\theta)},& \; \fEtrBS^\textbf{a}{}_i 	&= \diagmat{\fA (t,r),\fB (t,r) ,\fB (t,r)\sin (\theta)},& \\
	\gABS^i{}_j 	&= \diagmat{\gAo (t,r),\gAt (t,r) ,\gAt (t,r)},&  \fABS^i{}_j 	&= \diagmat{\fAo (t,r),\fAt (t,r) ,\fAt (t,r)},& \\
	\gL^i	 	&= \mat{\gLo (t,r) \\ 0 \\ 0}, \quad  \fL^i= \mat{\fLo (t,r) \\ 0 \\ 0},& \quad \hShift ^i &= \mat{\qo (t,r) \\ 0 \\ 0}, \quad \sLp^\textbf{a}= \mat{\po (t,r) \\ 0 \\ 0}. &
\end{alignat}
The metrics $\gSp$ and $\fSp$ share the background geometry, which is the spatial part of the flat metric in spherical coordinates,
\begin{align}
	\delta = \diagmat{1,r^2,r^2\sin(\theta)^2}.
\end{align}
All the fields depend on $t$ and $r$. We define \cite{doi:10.1063/1.5100027},
\begin{align}
	\mR 	&\coloneqq \dfrac{\ee^{2\fconf}\fB}{\ee^{2\gconf}\gB},
	\qquad
	\esp{n}{k}	\coloneqq \sum_{i=0}^n \binom{n}{i} \beta_{(i+k)} \mR^i, \quad \binom{n}{i} = \dfrac{n!}{i!(n-i)!},
\end{align}
where the $\beta_{(n)}$ are 5 dimensionless real constants appearing in the \change{bimetric potential \eqref{eq:bim-V}}.
The objects in the Lorentz frame read,
\begin{subequations}
\begin{alignat}{2}
\label{eq:Lorentzframe}
	\sRbarBS	&=\diagmat{\dfrac{\gA}{\fA}\sLt,\,\dfrac{\gB}{\fB},\,\dfrac{\gB}{\fB}}, \quad &\sRbar &= \dfrac{\ee^{2\gconf}}{\ee^{2\fconf}}\,\sRbarBS,  \\
	\sLs  &=\sLsBS= \diagmat{\sLt,1,1}, \quad &\sRs &=\sRsBS=\diagmat{1,1,1},
\end{alignat}
\end{subequations}
with $\sLp=(\po,0,0)$. The geometric mean is,
\begin{align}
\label{eq:geommean}
	\hShift	&= (\qo,0,0), \quad \hSpBS=\diagmat{\gA\,\fA\,\sLt,\, \gB\,\fB,\, \gB\,\fB}, \quad \hSp=\ee^{2(\gconf+\fconf)}\,\hSpBS.
\end{align}
Since $\sRs$ is the identity operator, its derivatives are identically zero. The computation of $\dt\sRs$ through the formula in \eqref{eq:dtsRs} consistently gives zero.

One can recast the equations presented in this appendix in the standard \threePlusOne formalism by imposing,
\begin{alignat}{4}
	\gKBS	 &=0, 	\qquad	& \fKBS &=0, \qquad & \gABS &=\gAo+2\gAt, \qquad & \quad\fABS &=\fAo+2\fAt, \nonumber \\
	\gABS_i		&=\gK _i,	& \fABS_i   &=\fK_i, & \gLo &=\biggrb{\dfrac{2r}{\gB^2}+\dfrac{\dr\gA}{\gA^3}-\dfrac{2\dr\gB}{\gA^2\gB}}, &\fLo &=\biggrb{\dfrac{2r}{\fB^2}+\dfrac{\dr\fA}{\fA^3}-\dfrac{2\dr\fB}{\fA^2\fB}}, \nonumber \\
	\gconf &=0, & \fconf &=0, & \dfrac{\pfg \gdet}{\gdet} &=-2\gLapse \Bigrb{\gK _1+2\gK _2}, & \dfrac{\pff \fdet}{\fdet} &= -2\fLapse \Bigrb{\fK _1+2\fK _2}.
\end{alignat}

It is worth noting that all the equations presented in this appendix are invariant under the exchange,
\begin{align}
	\beta_{(n)} \leftrightarrow \beta_{(4-n)}, \quad \gKappa \leftrightarrow \fKappa, \quad \sLp \rightarrow -\sLp, \quad X \leftrightarrow \widetilde{X}, \quad \widebar{X} \leftrightarrow \widehat{X},
\end{align}
where $X$ is any of the fields introduced above.

\paragraph{The mean standard gauge.}

We start by writing the \onePlusLog gauge condition \eqref{eq:onePLusLogmean} for the mean lapse, determining $\hK$ via \eqref{eq:hKdtchi}. Hence, we need the evolution equation for the conformal spatial part of the geometric mean,
\begin{subequations}
\begin{align}
\label{eq:dtchi}
   \pfh\hSpBS_{rr}  &=\fLapse\biggl[\dfrac{2}{3} \big(\fAt-\fAo\big)\gA\,\fA\,\sLt -\dfrac{\gA\,\partial_{r}\fA}{e^{2\fconf}\fA}\,\po\biggr] +\gLapse\biggl[\dfrac{2}{3} \big(\gAt-\gAo\big)\gA\,\fA\,\sLt +\dfrac{\fA\,\partial_{r}\gA}{e^{2\gconf}\gA}\,\po\biggr] \nonumber \\
   						&\quad +\dfrac{\fA\,\gA\,\po}{\sLt}\bigg(\dt\po-\qo\partial_{r}\po\bigg) +\dfrac{1}{6}\fA\,\gA\,\sLt \bigg(\dfrac{\pfg\gdet}{\gdet}+\dfrac{\pff\fdet}{\fdet} -12\,\partial_{r}\qo\bigg),\\
   \pfh\hSpBS_{\theta\theta}  &=\fLapse\biggl[\dfrac{1}{3} \big(\fAo -\fAt\big)\,\fB\,\gB -\dfrac{\gB\,\partial_{r}\fB}{e^{2\fconf}\fA\sLt}\po\biggr] +\gLapse\biggl[\dfrac{1}{3} \big(\gAo -\gAt\big)\,\gB\,\fB +\dfrac{\fB\,\partial_{r}\gB}{e^{2\gconf}\gA\sLt}\po\biggr] \nonumber \\
   						&\quad+\dfrac{1}{6}\fB\,\gB \bigg(\dfrac{\pff\fdet}{\fdet} +\dfrac{\pfg\gdet}{\gdet}\bigg),
\end{align}
\end{subequations}
where the time derivative of $\po$ is given by \cite{Torsello_2019},
\begin{align}
\label{eq:dtp}
	\esp{2}{1}\dt\po	&=\esp{2}{1}\qo\dr\po + \esp{2}{1}\sLt \rb{\dfrac{\dr\fLapse}{\ee^{2\fconf}\fA}-\dfrac{\dr\gLapse}{\ee^{2\gconf}\gA}} \nonumber \\
								&\quad +\gLapse\,\po\qb{\esp{2}{1}\rb{\gAo+\dfrac{1}{3}\gKBS}+2\rb{\gAt+\dfrac{1}{3}\gKBS}\esp{1}{1}} \nonumber\\
								 &\quad +\fLapse\,\po\qb{\esp{2}{1}\rb{\fAo+\dfrac{1}{3}\fKBS}+2\rb{\fAt+\dfrac{1}{3}\fKBS}\mR\esp{1}{2}} \nonumber\\
								&\quad +\dfrac{2}{\ee^{2\gconf}\gB}\rb{\gLapse\esp{1}{1}\dfrac{\dr\fB+2\,\fB\,\dr\fconf}{\fA}-\fLapse\esp{1}{2}\dfrac{\dr\gB+2\,\gB\,\dr\gconf}{\gA}} \nonumber \\
								&\quad +\dfrac{2\sLt}{\ee^{2\gconf}\gB}\rb{
								     \fLapse\esp{1}{2}\dfrac{\dr\fB+2\,\fB\,\dr\fconf}{\fA}-\gLapse\esp{1}{1}\dfrac{\dr\gB+2\,\gB\,\dr\gconf}{\gA}}.
\end{align}
The mean \onePlusLog gauge condition \eqref{eq:onePLusLogmean} reads,
\begin{align}
  & (\mW\sLt)^{1/2}\,\dt\hLapse  =\biggl[\po \biggr(\dfrac{1}{e^{2\fconf}\fA} -\dfrac{\mW}{e^{2\gconf}\gA}\biggl) +\qo (\mW\sLt)^{1/2}\biggl]\partial_{r}\hLapse +\dfrac{(\mW\sLt)^{1/2}}{2}\bigg[\dfrac{\pff\fdet}{\fdet} +\dfrac{\pfg\gdet}{\gdet}\bigg] \nonumber \\
   										&\quad- 2 (\mW\sLt)^{1/2}\partial_{r}\qo+\hLapse\Biggl\{(\po)^2\Biggl[\Biggr(\dfrac{1}{ e^{2\fconf}\fA} -\dfrac{\mW}{ e^{2\gconf}\gA}\Biggl)\dfrac{\partial_{r}\po}{2\sLt^2}   \nonumber \\
   										&\quad\quad\quad+\dfrac{1}{\sLt}\Bigg[\fAo +\dfrac{1}{3}\fKBS +2\mR \dfrac{\left\langle\mR\right\rangle^1_2}{\left\langle\mR\right\rangle^2_1}\Bigg(\fAt +\dfrac{\fKBS}{3}\Bigg)  +\mW\Bigg(\gAo + \dfrac{1}{3}\gKBS +\dfrac{2\left\langle\mR\right\rangle^1_1}{\left\langle\mR\right\rangle^2_1}\Bigg(\gAt+\dfrac{\gKBS}{3}\Bigg)\Bigg)\Biggl]\Biggl]   \nonumber \\
   										&\quad\quad+\po\Biggl[\dfrac{\mW}{\ee^{2\gconf}\gA}\Bigg( \dfrac{\partial_r\gA}{\gA}+\dfrac{2\,\partial_r\gB}{\gB} \Bigg)-\dfrac{1}{\ee^{2\fconf}\fA}\Bigg( \dfrac{\partial_r\fA}{\fA}+\dfrac{2\,\partial_r\fB}{\fB} \Bigg) - \Bigg(\dfrac{1}{ e^{2\gconf}\gA} +\dfrac{1}{ e^{2\fconf}\fA\mW}\Bigg)\dfrac{\partial_{r}\mW}{2} \nonumber \\
   										&\quad\quad\quad+\dfrac{2\mR}{e^{2\fconf}\fA}\Bigg(\dfrac{\partial_{r}\fB}{\fB}+2\,\partial_{r}\fconf \Bigg) \Bigg(\dfrac{\mW\left\langle\mR\right\rangle^1_1}{\sLt\left\langle\mR\right\rangle^2_1} +\dfrac{\left\langle\mR\right\rangle^1_2}{\left\langle\mR\right\rangle^2_1}\Bigg) \nonumber \\	
   										&\quad\quad\quad-\dfrac{2}{e^{2\gconf}\gA}\Bigg(\dfrac{\partial_{r}\gB}{\gB}+2\,\partial_{r}\gconf \Bigg)\Bigg(\dfrac{\mW\left\langle\mR\right\rangle^1_1}{\left\langle\mR\right\rangle^2_1} +\dfrac{\left\langle\mR\right\rangle^1_2}{\sLt\left\langle\mR\right\rangle^2_1}\Bigg)  \Biggl]\Biggl\}.
\end{align}
We choose the flat metric in spherical polar coordinates as the background metric for the mean sector, so the $\Gamma$-driver gauge condition for the mean shift vector reads,
\begin{subequations}
\begin{align}
   \dt\qo  &=\dfrac{1}{4}\bigl(3\hShiftB + 4\qo\,\partial_r\qo\bigl), \\
   \partial_t \hShiftB^r 	&= \qo \partial_r\hShiftB^r +\bigrb{\partial_t\hL^r -\qo \partial_r\hL^r} -\eta \hShiftB^r.\label{eq:Gdriver2}
\end{align}
\end{subequations}
We remark that one can choose any connection as the background connection for the mean sector. For example, setting the mean connection $\big(\gGBS^i_{jk} +\fGBS^i_{jk}\big)/2$ as the background connection, keeps the symmetry between the metric sectors. The mean connection is,
\begin{subequations}
\begin{align}
   \dfrac{1}{2}\Big(\gGBS^1_{jk} +\fGBS^1_{jk}\Big)  &=\dfrac{1}{2}\mat{\dfrac{\partial_{r}\fA}{\fA} +\dfrac{\partial_{r}\gA}{\gA} & 0 & 0\\ 0 & -\dfrac{\fB\,\partial_{r}\fB}{\fA^2} -\dfrac{\gB\,\partial_{r}\gB}{\gA^2} & 0\\ 0 & 0 &-\bigg(\dfrac{\fB\,\partial_{r}\fB}{\fA^2} +\dfrac{\gB\,\partial_{r}\gB}{\gA^2}\bigg)\sin(\theta)^2}, \\
   \dfrac{1}{2}\Big(\gGBS^2_{jk} +\fGBS^2_{jk}\Big)  &=\dfrac{1}{2}\mat{0 &\dfrac{\partial_{r}\fB}{\fB} +\dfrac{\partial_{r}\gB}{\gB} & 0\\\dfrac{\partial_{r}\fB}{\fB} +\dfrac{\partial_{r}\gB}{\gB} & 0 & 0\\ 0 & 0 & -2\cos(\theta)\sin(\theta)}, \\
   \dfrac{1}{2}\Big(\gGBS^3_{jk} +\fGBS^3_{jk}\Big)  &=\dfrac{1}{2}\mat{0 & 0 &\dfrac{\partial_{r}\fB}{\fB} +\dfrac{\partial_{r}\gB}{\gB}\\ 0 & 0 &2\cot(\theta)\\\dfrac{\partial_{r}\fB}{\fB} +\dfrac{\partial_{r}\gB}{\gB} &2\cot(\theta) & 0}.
\end{align}
\end{subequations}
In order to write $\dt\hL^r$ in \eqref{eq:Gdriver2}, we need the evolution equation of the Christoffel symbol of the conformal mean metric, contracted with the inverse conformal mean metric; this is,
\begin{align}
\label{eq:dtG}
   \hSpBS^{jk}\pfh\hGBS^r_{jk}  &=\dfrac{\pfh\hSpBS^{\theta\theta}}{\fA\,\gA\,\sLt}  \bigg(\gB\,\partial_{r}\fB +\fB\,\partial_{r}\gB\bigg)+\dfrac{\pfh\hSpBS_{\theta\theta}}{\fA\,\fB\,\gA\,\gB\,\sLt} \bigg(\dfrac{\partial_{r}\fB}{\fB} +\dfrac{\partial_{r}\gB}{\gB}\bigg) -\dfrac{\partial_{r}\pfh\hSpBS_{rr}}{2\fA^2\gA^2\sLt^2} -\dfrac{\partial_{r}\pfh\hSpBS_{\theta\theta}}{\fA\,\fB\,\gA\,\gB\,\sLt} -\partial_{r}\pfh\hSpBS^{rr} \nonumber \\
   											&\quad+\bigg(\dfrac{\pfh\hSpBS_{rr}}{2\fA^2\gA^2\sLt^2}-\pfh\hSpBS^{rr}e^{-2(\fconf +\gconf)}\bigg) \bigg(\dfrac{\partial_{r}\fA}{\fA} +\dfrac{\partial_{r}\gA}{\gA} +\dfrac{\partial_{r}\sLt}{\sLt}\bigg) -\pfh\hSpBS^{rr} \bigg(\dfrac{\partial_{r}\fB}{\fB} +\dfrac{\partial_{r}\gB}{\gB}\bigg).
\end{align}
The evolution equation for the mean conformal connection $\hL^r$ is,
\begin{align}
\label{eq:dtL}
   \dt\hL^{r}  &=\hSpBS^{jk}\pfh\hGBS^r_{jk} +\dfrac{\pfh\hSpBS^{rr}}{2} \bigg(\dfrac{\partial_{r}\fA}{\fA} +\dfrac{\partial_{r}\gA}{\gA} +\dfrac{\po\partial_{r}\po}{\sLt^2}\bigg) -\dfrac{\po\partial_{r}\po\partial_{r}\qo}{2\,\fA\,\gA\,\sLt^3} -\dfrac{\partial_{rr}\qo}{\fA\,\gA\,\sLt} \nonumber \\
   					&\quad+\pfh\hSpBS^{\theta\theta} \bigg( 2 r-\dfrac{\gB\,\partial_{r}\fB}{\fA\,\gA\,\sLt} -\dfrac{\fB\,\partial_{r}\gB}{\fA\,\gA\,\sLt} \bigg) +\dfrac{\partial_{r}\qo}{\gA\,\fA\,\sLt} \bigg(\dfrac{\partial_{r}\gB}{\gB}-\dfrac{\partial_{r}\gA}{2\gA}+\dfrac{\partial_{r}\fB}{\fB} -\dfrac{\partial_{r}\fA}{2\fA} -\dfrac{4\,\fA\,\gA\,\sLt\, r}{\fB\,\gB}\bigg) \nonumber \\
   					&\quad+\dfrac{\qo}{\gA\,\fA\,\sLt}\biggl\{\dfrac{4\gA\,\fA\,\sLt}{\fB\,\gB} +\partial_r\biggrb{\dfrac{\partial_r\fA}{2\fA}-\dfrac{\partial_r\fB}{\fB}} +\partial_r\biggrb{\dfrac{\partial_r\gA}{2\gA}-\dfrac{\partial_r\gB}{\gB}} +\partial_r\biggrb{\dfrac{\partial_r\sLt}{2\sLt}}  \nonumber \\
   					&\quad\quad+\bigg[\dfrac{\partial_{r}\gB}{\gB}+\dfrac{\partial_{r}\fB}{\fB}-\dfrac{1}{2}\bigg(\dfrac{\partial_{r}\fA}{\fA} +\dfrac{\partial_{r}\gA}{\gA}\bigg)\bigg] \bigg(\dfrac{\partial_{r}\fA}{\fA} +\dfrac{\partial_{r}\gA}{\gA}\bigg) - 2\bigg(\dfrac{\partial_{r}\fB}{\fB} +\dfrac{\partial_{r}\gB}{\gB}\bigg) \dfrac{\gA\,\fA}{\gB\,\fB}\sLt \, r \nonumber \\
   					&\quad\quad+ \bigg(\dfrac{\partial_{r}\fB}{\fB} -\dfrac{\partial_{r}\fA}{\fA} -\dfrac{\partial_{r}\gA}{\gA} +\dfrac{\partial_{r}\gB}{\gB}\bigg)\dfrac{\po\partial_{r}\po}{\sLt^2} - \po^2\dfrac{(\partial_{r}\po)^2}{2\sLt^4} \biggl\}.
\end{align}
Finally, the evolution equation \eqref{eq:Gdriver2} for the variable $\hShiftB$ becomes,
\begin{subequations}
\begin{align}
   \dt\hShiftB  &=\dt\hL^{r} -\eta\hShiftB +\qo \partial_r\hShiftB  \nonumber \\
   					&\quad-\dfrac{\qo}{\gA\,\fA\,\sLt}\biggl\{\dfrac{2\gA\,\fA\,\sLt}{\fB\,\gB} +\partial_r\biggrb{\dfrac{\partial_r\fA}{2\fA}-\dfrac{\partial_r\fB}{\fB}} +\partial_r\biggrb{\dfrac{\partial_r\gA}{2\gA}-\dfrac{\partial_r\gB}{\gB}} +\partial_r\biggrb{\dfrac{\partial_r\sLt}{2\sLt}}  \nonumber \\
   					&\quad\quad+\bigg[\dfrac{\partial_{r}\gB}{\gB}+\dfrac{\partial_{r}\fB}{\fB}-\dfrac{1}{2}\bigg(\dfrac{\partial_{r}\fA}{\fA} +\dfrac{\partial_{r}\gA}{\gA}\bigg)\bigg] \bigg(\dfrac{\partial_{r}\fA}{\fA} +\dfrac{\partial_{r}\gA}{\gA}\bigg) - 2\bigg(\dfrac{\partial_{r}\fB}{\fB} +\dfrac{\partial_{r}\gB}{\gB}\bigg) \dfrac{\gA\,\fA}{\gB\,\fB}\sLt \, r \nonumber \\
   					&\quad\quad+ \bigg(\dfrac{\partial_{r}\fB}{\fB} -\dfrac{\partial_{r}\fA}{\fA} -\dfrac{\partial_{r}\gA}{\gA} +\dfrac{\partial_{r}\gB}{\gB}\bigg)\dfrac{\po\partial_{r}\po}{\sLt^2} - \po^2\dfrac{(\partial_{r}\po)^2}{2\sLt^4} \biggl\},\\
   					&=-\eta\hShiftB +\qo \bigg(\partial_r\hShiftB +\dfrac{2}{\gA\fA} \bigg) +\hSpBS^{jk}\pfh\hGBS^r_{jk} +\dfrac{\pfh\hSpBS^{rr}}{2} \bigg(\dfrac{\partial_{r}\fA}{\fA} +\dfrac{\partial_{r}\gA}{\gA} +\dfrac{\po\partial_{r}\po}{\sLt^2}\bigg) -\dfrac{\po\partial_{r}\po\partial_{r}\qo}{2\,\fA\,\gA\,\sLt^3} \nonumber\\
   					&\quad-\dfrac{\partial_{rr}\qo}{\fA\,\gA\,\sLt}+\pfh\hSpBS^{\theta\theta} \bigg( 2 r-\dfrac{\gB\,\partial_{r}\fB}{\fA\,\gA\,\sLt} -\dfrac{\fB\,\partial_{r}\gB}{\fA\,\gA\,\sLt} \bigg) +\dfrac{\partial_{r}\qo}{\gA\,\fA\,\sLt} \bigg(\dfrac{\partial_{r}\gB}{\gB}-\dfrac{\partial_{r}\gA}{2\gA}+\dfrac{\partial_{r}\fB}{\fB} -\dfrac{\partial_{r}\fA}{2\fA} -\dfrac{4\,\fA\,\gA\,\sLt\, r}{\fB\,\gB}\bigg)
\end{align}
\end{subequations}

\paragraph{The mean maximal slicing.} In spherical symmetry, the trace $\hK$ of the extrinsic curvature of the spatial part $\hSp$ of the geometric mean metric $\hMet$ is,
\begin{align}
\label{eq:hKapp}
	\hK	&=\dfrac{1}{2}\Bigrb{\mV^r\partial_r\log(\hLapse)+\mF}.
\end{align}
The two functions $\mV^r$ and $\mF$ are given by,
\begin{subequations}
\begin{align}
   \sqrt{\mW\sLt}\,\mV^r  &=\bigg(\dfrac{\mW}{e^{2\gconf}\gA} -\dfrac{1}{e^{2\fconf}\fA}\bigg)\bigg(\dfrac{\sLt-1}{2\sLt}\bigg)\po, \\
   \esp{2}{1}\sqrt{\mW\sLt}\,\mF 	&=-\bigg[\bigg(\gAt +\dfrac{\gKBS}{3}\bigg)\mW\left\langle\mR\right\rangle^1_1 +\bigg(\fAt +\dfrac{\fKBS}{3}\bigg)\mR\left\langle\mR\right\rangle^1_2\bigg]\dfrac{\po^2}{\sLt} \nonumber \\
   													&\quad+\bigg[\mW\left\langle\mR\right\rangle^1_1\biggl(\dfrac{2\,\gB\,\partial_{r}\gconf +\partial_{r}\gB}{\gA}-\dfrac{2\,\fB\,\partial_{r}\fconf +\partial_{r}\fB}{\fA\sLt}\biggr) \nonumber \\
   													&\quad\quad+\left\langle\mR\right\rangle^1_2\bigg(\dfrac{2\,\gB\,\partial_{r}\gconf +\partial_{r}\gB}{\gA\sLt}-\dfrac{2\,\fB\,\partial_{r}\fconf +\partial_{r}\fB}{\fA}\bigg)\bigg]\dfrac{\po}{\ee^{2\gconf}\gB} \nonumber \\
   													&\quad+\biggl\{\dfrac{1}{2}\bigg(\fKBS +\mW\gKBS\bigg)\sLt + \bigg(\dfrac{1}{e^{2\fconf}\fA} -\dfrac{\mW}{e^{2\gconf}\gA}\bigg)\dfrac{\partial_{r}\po}{2\sLt} \nonumber \\
   													&\quad\quad+\bigg[\bigg(\dfrac{\mW}{e^{2\gconf}\gA}-\dfrac{1}{e^{2\fconf}\fA}\bigg) \big(3+\sLt\big)\dfrac{\partial_{r}\po}{4\sLt}-\dfrac{\sLt}{2}\bigg(\mW\bigg(\gAo +\dfrac{\gKBS}{3}\bigg)+\fAo +\dfrac{\fKBS}{3}\bigg)\bigg]\dfrac{\po^2}{\sLt^2} \nonumber \\
   													&\quad\quad+\po\bigg[\dfrac{1}{\ee^{2\fconf}\fA\sLt}\bigg(\dfrac{\partial_{r}\fdet}{4\fdet} -\partial_{r}\fconf -\dfrac{\partial_{r}\fA}{2\fA}\bigg) -\dfrac{\mW}{\ee^{2\gconf}\gA\sLt} \bigg(\dfrac{\partial_{r}\gdet}{4\gdet} -\partial_{r}\gconf -\dfrac{\partial_{r}\gA}{2\gA}\bigg) \nonumber \\
   													&\quad\quad\quad+ \bigg(\dfrac{1}{e^{2\fconf}\fA\mW}+\dfrac{1}{e^{2\gconf}\gA}\bigg)\bigg( \dfrac{\sLt-1}{\sLt}\bigg)\dfrac{\partial_{r}\mW}{4}\bigg]\biggr\}\left\langle\mR\right\rangle^2_1.
\end{align}
\end{subequations}
The boundary value problem \eqref{eq:meanmaxfinal2} contains the time derivatives of $\mV^r$ and $\mF$. These are quite lengthy, so we do not write them down explicitly here.

\bibliographystyle{JHEP}
\bibliography{biblio}

\end{document}

%% file: macros.tex
\usepackage{mathabx}
\usepackage{accents}
\usepackage{xspace}

\DeclareMathAlphabet\urwscr{U}{urwchancal}{m}{n}%

\ifx \ii \undefined  % Avoid the redefinitions
\DeclareMathAlphabet{\mathsfit}{\encodingdefault}{\sfdefault}{m}{sl}
\SetMathAlphabet{\mathsfit}{bold}{\encodingdefault}{\sfdefault}{bx}{sl}

\DeclareSymbolFont{eulerletters}{U}{eus}{m}{n}
\DeclareMathSymbol{\eulU}{\mathord}{eulerletters}{`U}
\DeclareMathSymbol{\eulC}{\mathord}{eulerletters}{`C}
\DeclareMathSymbol{\eulB}{\mathord}{eulerletters}{`B}
\DeclareMathSymbol{\eulQ}{\mathord}{eulerletters}{`Q}
\DeclareMathSymbol{\eulE}{\mathord}{eulerletters}{`E}
\DeclareMathSymbol{\eulD}{\mathord}{eulerletters}{`D}
\DeclareMathSymbol{\eulV}{\mathord}{eulerletters}{`V}
\DeclareMathSymbol{\eulW}{\mathord}{eulerletters}{`W}
\DeclareMathSymbol{\eulS}{\mathord}{eulerletters}{`S}
\DeclareMathSymbol{\eulG}{\mathord}{eulerletters}{`G}
\DeclareMathSymbol{\eulL}{\mathord}{eulerletters}{`L}
\DeclareMathSymbol{\eulK}{\mathord}{eulerletters}{`K}
\DeclareMathSymbol{\eulR}{\mathord}{eulerletters}{`R}
\DeclareMathSymbol{\eulN}{\mathord}{eulerletters}{`N}

\DeclareSymbolFont{CMB}{OMS}{cmsy}{m}{it}
\DeclareMathSymbol{\cmA}{\mathord}{CMB}{`A}
\DeclareMathSymbol{\cmB}{\mathord}{CMB}{`B}
\DeclareMathSymbol{\cmW}{\mathord}{CMB}{`W}

\definecolor{red}{rgb}{1,0,0.1}
\definecolor{green}{rgb}{0.0,0.6,0}
\definecolor{blue}{rgb}{0.1,0.1,1}
\definecolor{orange}{rgb}{0.6,0.3,0}
\definecolor{magenta}{rgb}{0.9,0.1,1}

\newcommand{\qm}[1]{``#1"}
\newcommand{\vect}[1]{\operatorname{vec}\qb{#1}}

\newcommand\nPlusOne{$\sdim\hspace{-0.1em}+\hspace{-0.1em}1$\xspace}
\newcommand\threePlusOne{$3\hspace{-0.075em}+\hspace{-0.1em}1$\xspace}
\newcommand\onePlusLog{1+$\hspace{0.1em}\log$\xspace}

\renewcommand\tilde[1]{\mkern1mu\widetilde{\mkern-1mu#1}}

\global\long\def\dt{\partial_{t}}
\global\long\def\dtp{\partial_{t} \sLp}

\global\long\def\bigqb#1{\bigl[#1\bigr]}
\global\long\def\Bigqb#1{\Bigl[#1\Bigr]}
\global\long\def\biggqb#1{\biggl[#1\biggr]}

\global\long\def\bigrb#1{\bigl(#1\bigr)}
\global\long\def\Bigrb#1{\Bigl(#1\Bigr)}
\global\long\def\biggrb#1{\biggl(#1\biggr)}
\global\long\def\rb#1{\left(#1\right)}
 \global\long\def\qb#1{\left[#1\right]}
 \global\long\def\cb#1{\left\lbrace #1 \right\rbrace }
 \global\long\def\qm#1{``#1''}
 
 \global\long\def\brb#1{\Bigl(#1\Bigr)}

\global\long\def\vect#1{\operatorname{vec}\qb{#1}}

\newcommand\bSe{\begin{subequations}}
\newcommand\eSe{\end{subequations}}

\newcommand\mat[1]{
\begin{pmatrix}
#1
\end{pmatrix}
}

\newcommand\diagmat[1]{\operatorname{diag}\qb{#1}}

\usepackage{xparse}
\ExplSyntaxOn
\NewDocumentCommand{\mref}{m}{\quinn_mref:n {#1}}
\seq_new:N \l_quinn_mref_seq
\cs_new:Npn \quinn_mref:n #1
 {
  \seq_set_split:Nnn \l_quinn_mref_seq { , } { #1 }
  \seq_pop_right:NN \l_quinn_mref_seq \l_tmpa_tl
  ( % print the left parenthesis
  \seq_map_inline:Nn \l_quinn_mref_seq
    { \ref{##1},\nobreakspace } % print the first references
  \exp_args:NV \ref \l_tmpa_tl % print the last or only one
  ) % print the right parenthesis
 }
\ExplSyntaxOff

\newcommand\ii{\mathrm{i}}
\newcommand\ee{\mathrm{e}}
\newcommand\dd{\mathrm{d}}

\newcommand\tr{\mathsf{{\scriptscriptstyle T}}}

\newcommand\op[1]{\operatorname{#1}}

\newcommand\dr{\partial _r}

\newcommand\sdim{N}

\ifColors
  \newcommand\gSector[1]{{\color{blue}#1}}
  \newcommand\fSector[1]{{\color{red}#1}}
  \newcommand\hSector[1]{{\color{green}#1}}
  \newcommand\lSector[1]{{\color{BurntOrange}#1}}
  \newcommand\mSector[1]{{\color{magenta}#1}}

\else
  \newcommand\gSector[1]{{\color{black}#1}}
  \newcommand\fSector[1]{{\color{black}#1}}
  \newcommand\hSector[1]{{\color{black}#1}}
  \newcommand\lSector[1]{{\color{black}#1}}
  \newcommand\mSector[1]{{\color{black}#1}}

\fi

\newcommand\hashsize{4.25pt}

\newcommand\gcauchy{\gSector{\theta_g}}
\newcommand\fcauchy{\fSector{\theta_f}}

\newcommand\gEm{\gSector{\mathcal{E}_g}}
\newcommand\fEm{\fSector{\mathcal{E}_f}}

\newcommand\spincon{\scaleto{\eulW}{5.5pt}}
\newcommand\gspincon{\gSector{\scaleto{\eulW}{5.5pt}_g}}

\newcommand\gcauchycon{\gSector{\omega_g}}

\newcommand\gMet{\gSector g}
\newcommand\gSp{\gSector{\gamma}}
\newcommand\gLapse{\gSector{\alpha}}
\newcommand\gShift{\gSector \beta}

\newcommand\gK{\gSector K}
\newcommand\gE{\gSector e}
\newcommand\gFE{\gSector E}
\newcommand\gD{\gSector{D}}

\newcommand\bimV{\mSector{V}}
\newcommand\gLmat{\gSector{\mathcal{L}_{g}}}
\newcommand\fLmat{\fSector{\mathcal{L}_{f}}}

\newcommand\gRicci{\gSector{R_{g}}}

\newcommand\gCD{\gSector{\nabla}}

\newcommand\hCD{\hSector{\accentset{\scaleto{\#}{\hashsize}}{\nabla}}}

\newcommand\gnor{\gSector{\nu}}
\newcommand\fnor{\fSector{\tilde{\nu}}}
\newcommand\hnor{\hSector{\accentset{\scaleto{\#}{\hashsize}}{\nu}}}

\newcommand\fMet{\fSector f}
\newcommand\fSp{\fSector{\varphi}}
\newcommand\fLapse{\fSector{\tilde{\alpha}}}
\newcommand\fShift{\fSector{\tilde{\beta}}}

\newcommand\fK{\fSector{\tilde{K}}}
\newcommand\fE{\fSector m}

\newcommand\fEtr{\fSector{m_\mathsf{o}}}

\newcommand\fFEtr{\fSector{M_\mathsf{o}}}
\newcommand\fD{\fSector{\tilde{D}}}

\newcommand\fRicci{\fSector{R_{f}}}

\newcommand\fCD{\fSector{\widetilde{\nabla}}}

\newcommand\mSqrt{\mSector{S}}

\newcommand\mW{\mSector{W}}

\newcommand\mF{\mSector{\mathcal{M}}}
\newcommand\mV{\mSector{\mathcal{N}}}
\newcommand\mP{\mSector{\mathcal{P}}}

\newcommand\hh{\hSector{\xi}}%\mathcal{h}
\newcommand\heta{\hSector{\zeta}}

\newcommand\gKappa{\gSector{\kappa_{g}}}

\newcommand\fKappa{\fSector{\kappa_{f}}}

\newcommand\gjotab{\gSector{j^\mathrm{b}}}

\newcommand\grhoeff{\gSector{\rho^\mathrm{eff}}}
\newcommand\gjotaeff{\gSector{j^\mathrm{eff}}}
\newcommand\gJotaeff{\gSector{J^\mathrm{eff}}}

\newcommand\frhoeff{\fSector{\tilde{\rho}{}^\mathrm{eff}}}

\newcommand\fJotaeff{\fSector{\tilde{J}{}^\mathrm{eff}}}

\newcommand\sgn{\gSector{\mathsfit{n}{\mkern1mu}}}

\newcommand\sgB{\gSector{\eulB}}

\newcommand\sfn{\fSector{\tilde{\mathsfit{n}}{\mkern1mu}}}

\newcommand\hK{\hSector{\accentset{\scaleto{\#}{\hashsize}}{K}}}
\newcommand\hG{\hSector{\accentset{\scaleto{\#}{\hashsize}}{\Gamma}}}

\newcommand\gBSSN[1]{{\widebar{#1}}}
\newcommand\fBSSN[1]{{\widehat{#1}}}
\newcommand\hBSSN[1]{{\accentset{\circ}{{#1}}}}
\newcommand\lBSSN[1]{{\accentset{\ast}{{#1}}}}

\newcommand\hSpBS{\hSector{\hBSSN \chi}}

\newcommand\pfg{\gSector{\partial_{\perp}}}
\newcommand\pff{\fSector{\widetilde{\partial}_{\perp}}}
\newcommand\pfh{\hSector{\partial_{\scaleto{\#}{\hashsize}}}}
\newcommand\gconf{\gSector{\phi}}
\newcommand\fconf{\fSector{\psi}}

\newcommand\gSpBS{\gSector{\gBSSN \gamma}}

\newcommand\gShiftB{\gSector B}

\newcommand\gABS{\gSector{\gBSSN A}}
\newcommand\gKBS{\gSector{\gBSSN K}}
\newcommand\gEBS{\gSector{\gBSSN e}}
\newcommand\gDBS{\gSector{\gBSSN{D}}}

\newcommand\gG{\gSector{\Gamma}}
\newcommand\gGBS{\gSector{\gBSSN \Gamma}}
\newcommand\gL{\gSector{\gBSSN \Lambda}}
\newcommand\gDG{\gSector{\bigtriangleup\gBSSN{\Gamma}}}
\newcommand\gDback{\gSector{\gBSSN{D}_\mathsf{B}}}
\newcommand\gGbackBS{\gSector{\gBSSN{\Gamma}_\mathsf{B}}}
\newcommand\gRback{\gSector{\gBSSN{R}_\mathsf{B}}}

\newcommand\gA{\gSector{\gBSSN{a}}}
\newcommand\gB{\gSector{\gBSSN{b}}}
\newcommand\gAo{\gSector{\gBSSN{A}_1}}
\newcommand\gAt{\gSector{\gBSSN{A}_2}}
\newcommand\gLo{\gSector{\gBSSN{\Lambda}^1}}

\newcommand\po{\lSector{\sLp^1}}

\newcommand\qo{\hSector{\hShift^r}}

\newcommand\fSpBS{\fSector{\fBSSN \varphi}}

\newcommand\fABS{\fSector{\fBSSN A}}
\newcommand\fKBS{\fSector{\fBSSN K}}

\newcommand\fEtrBS{\fSector{\fBSSN{m}_\mathrm{o}}}

\newcommand\fG{\fSector{\tilde \Gamma}}
\newcommand\fGBS{\fSector{\fBSSN \Gamma}}
\newcommand\fL{\fSector{\fBSSN \Lambda}}

\newcommand\fA{\fSector{\fBSSN{a}}}
\newcommand\fB{\fSector{\fBSSN{b}}}
\newcommand\fAo{\fSector{\fBSSN{A}_1}}
\newcommand\fAt{\fSector{\fBSSN{A}_2}}
\newcommand\fLo{\fSector{\fBSSN{\Lambda}^1}}

\newcommand\hDBS{\hSector{\hBSSN{D}}}
\newcommand\hGBS{\hSector{\hBSSN \Gamma}}
\newcommand\hDback{\hSector{\hBSSN{D}_\mathsf{B}}}

\newcommand\hShiftB{\hSector{\hBSSN B}}
\newcommand\hL{\hSector{\hBSSN \Lambda}}
\newcommand\hDG{\hSector{\bigtriangleup\hBSSN{\Gamma}}}
\newcommand\hGbackBS{\hSector{\hBSSN{\Gamma}_\mathsf{B}}}
\newcommand\hRback{\hSector{\hBSSN{R}_\mathsf{B}}}
\newcommand\hRiBS{\hSector{\hBSSN{R}}}

\newcommand\gdetBS{\gSector{\gBSSN \Delta}}
\newcommand\fdetBS{\fSector{\fBSSN \Delta}}
\newcommand\hdetBS{\hSector{\accentset{\circ}{\Delta}}}

\newcommand\sgBBS{\gSector{\gBSSN \eulB}}

\newcommand\sLsBS{\lSector{\boldsymbol{\lBSSN{\Lambda}}_\mathrm{s}}}
\newcommand\sRsBS{\lSector{\boldsymbol{\lBSSN{\mathrm{R}}}}}
\newcommand\sRbarBS{\lSector{\boldsymbol{\lBSSN{\mathrm{R}}_\mathrm{o}}}}

\newcommand\hMet{\hSector h}
\newcommand\hSp{\hSector{\chi}}
\newcommand\hLapse{\hSector H}
\newcommand\hShift{\hSector q}

\newcommand\sLs{\lSector{\boldsymbol{\Lambda}_\mathrm{s}}}
\newcommand\sLt{\lSector{\lambda}}

\newcommand\sLp{\lSector{\boldsymbol{\mathrm{p}}}}
\newcommand\sRs{\lSector{\boldsymbol{\mathrm{R}}}}
\newcommand\sRbar{\lSector{\boldsymbol{\mathrm{R}}_\mathrm{o}}}

\newcommand\sI{\lSector{\boldsymbol{\mathrm{I}}}}
\newcommand\sEta{\lSector{\boldsymbol{\mathrm{\delta}}}}
\newcommand\Eta{\lSector{\boldsymbol{\mathrm{\eta}}}}

\newcommand\gdet{\gSector{\Delta}}
\newcommand\fdet{\fSector{\tilde \Delta}}
\newcommand\hdet{\hSector{\accentset{\scaleto{\#}{\hashsize}}{\Delta}}}

\newcommand\mR{\mSector{\wideparen{R}}}
\newcommand\esp[2]{\left\langle\mR\right\rangle ^{#1}_{#2}}

\ifColors

\else

\fi

\fi  % avoid the redefinitions